\documentclass[twocolumn]{aa}
\usepackage[colorlinks,linkcolor={blue},citecolor={blue},urlcolor={red}]{hyperref}
\usepackage{graphicx}
\usepackage{url}
\usepackage{natbib}
\usepackage{xspace}

\def\apjs{ApJS}

\def\beq{\begin{equation}}
\def\eeq{\end{equation}}
\def\bey{\begin{eqnarray}}
\def\eey{\end{eqnarray}}

\def\Mpc{\,{\rm Mpc}}
\def\mpc{\, h^{-1}{\rm {Mpc}}}

\def\kms{\,{\rm {km\, s^{-1}}}}
\def\Msun{{\rm M_\odot}}
\def\msun{\, h^{-1}{\rm M_\odot}}

\def\gs{\mathrel{\raise1.16pt\hbox{$>$}\kern-7.0pt
\lower3.06pt\hbox{{$\scriptstyle \sim$}}}}
\def\ls{\mathrel{\raise1.16pt\hbox{$<$}\kern-7.0pt
\lower3.06pt\hbox{{$\scriptstyle \sim$}}}}
\def\gtsima{\, {\buildrel > \over \sim} \,}
\def\ltsima{\, {\buildrel < \over \sim} \,}
\def\prosima{\, {\buildrel \propto \over \sim} \,}
\def\gsim{\lower.5ex\hbox{\gtsima}}
\def\lsim{\lower.5ex\hbox{\ltsima}}
\def\simgt{\lower.5ex\hbox{\gtsima}}
\def\simlt{\lower.5ex\hbox{\ltsima}}
\def\simpr{\lower.5ex\hbox{\prosima}}

\def\vr{v_{\rm max}/v_{\rm 200}}
\def\zf{z_{\rm f}}
\begin{document}

\title{Evaluating the origins of the secondary bias based on the correlation of halo properties with the linear density field}
\titlerunning{Secondary bias in linear density field}
\authorrunning{X.Y. Wang et al.}
\author{Xiaoyu Wang\inst{1,2}, Huiyuan Wang\inst{1,2}, H.J. Mo\inst{3}, JingJing Shi\inst{4}, Yipeng Jing\inst{5,1}}
\institute{Key Laboratory for Research in Galaxies and Cosmology, Department of Astronomy, University of Science and Technology of China, Hefei, Anhui 230026, China; wambyybz@mail.ustc.edu.cn, whywang@ustc.edu.cn
\and School of Astronomy and Space Science, University of Science and Technology of China, Hefei 230026, China
\and Department of Astronomy, University of Massachusetts, Amherst MA 01003-9305, USA
\and Institute for the Physics and Mathematics of the Universe (Kavli IPMU, WPI), UTIAS, Tokyo Institutes for Advanced Study, University of Tokyo, Chiba, 277-8583, Japan
\and Department of Astronomy, and IFSA Collaborative Innovation Center, Shanghai Jiao Tong University, Shanghai 200240, China}

\abstract{
Using two sets of large $N$-body simulations, we study the origin of
the correlations of halo assembly time ($\zf$), concentration 
($\vr$) and spin ($\lambda$) with the large-scale evolved density 
field at given halo mass, i.e. the secondary bias. We find that 
the secondary bias is the secondary effect of the correlations 
of halo properties with the linear density estimated at the same comoving scale. 
Using the linear density on different scales, 
we find two types of correlations. The internal correlation, 
which reflects the correlation of halo properties with 
the mean linear over-density $\delta_{\rm L}$
within the halo Lagrangian radius $R_{\rm L}$, is positive 
for both $\zf$ and $\vr$, and negative for $\lambda$. 
The external correlation, which describes the correlation of halo 
properties with linear overdensity at $R>R_{\rm L}$ for given  
$\delta_{\rm L}$, shows trends opposite 
to the internal correlation. Both of the external and internal correlations 
depend only weakly on halo mass, indicating a similar origin 
for halos of different masses. 
Our findings offer a transparent perspective on the origin of 
the secondary bias. The secondary bias can be 
largely explained by the competition of the external and
internal correlations together with the correlation of the 
linear density field on different scales. These two types of correlations 
combined can establish the complex halo-mass dependence 
of the secondary bias observed in the simulations.
}

\maketitle
\section{Introduction}
\label{sec_intro}

Numerical $N$-body simulations have revealed that the clustering 
of dark matter halos depends not only on halo mass 
\citep[e.g.][]{Mo1996, Sheth2001} but also on other halo properties, 
such as mass assembly history, and structural and dynamical 
properties \citep{Gao05, Wechsler06, Wetzel07, Jing07, Bett07, Gao07, LiY08, Faltenbacher2010, Lacerna11, Lazeyras17, xu18, Salcedo2018, Mao18, Chue2018}. 
The dependence of the halo clustering on halo properties 
other than the halo mass is usually referred to as the halo assembly bias 
or the secondary bias. Understanding such secondary bias of the halo population is 
important not only for understanding the formation of dark matter 
halos in the cosmic density field, but also for understanding 
galaxy formation and evolution in dark matter halos 
\citep[see e.g.][]{Zhu06, Yang06, Zentner14, Hearin15, Wechsler18, WangH2018, Salcedo2020}. 

It is well known that halo properties, such as 
assembly time, concentration, substructure and spin, 
are correlated among themselves 
\citep[e.g.][]{Jing02, Zhao03, Gao04, Allgood06, Hahn07, WangH11, Mao18}. 
However, these properties exhibit complex trends in their secondary bias,
sometimes in a way different from that expected from their mutual correlations. 
For instance, the dependence of the halo clustering 
on the assembly time is usually strong for low-mass halos, 
but weak at the massive end \citep[see e.g. ][]{Gao05, Jing07, LiY08, Mao18, Chue2018},
while the secondary bias for the halo spin parameter increases 
with halo mass \citep[e.g. ][]{Bett07, Gao07, Faltenbacher2010, Salcedo2018}.
Moreover, the dependence of the secondary bias on the two halo 
structural parameters, halo concentration and subhalo 
abundance, changes sign at around the characteristic 
mass of collapse
\citep[][]{Wechsler06, Gao07, Salcedo2018}. 

These results imply that the secondary bias has multiple origins. 
This is supported by numerous investigations in the literature, 
most of which focused on the secondary bias in the assembly 
time. For example, \cite{WangH07} found that old 
small halos are usually located closer to massive 
structures than their younger counterparts \citep[see also][]{Hahn09}, 
a phenomenon referred to as the neighbour bias by \cite{Salcedo2018}. 
Further studies showed that the secondary bias may also be 
related to the nearby cosmic web of halos 
\citep{Yang17, Paranjape2018, Ramakrishnan19}.  
Several processes related to the presence of massive neighbours 
have been proposed. For instance, the tidal field of the massive 
structure can accelerate the ambient matter and truncate 
the mass accretion onto small halos 
\citep{WangH07, Hahn09, WangH11, Shi15, Paranjape2018, Mansfield20}. 
Splashback halos that have ever passed through massive 
host halos may be severely stripped by the tidal force of 
the hosts \citep{Ludlow09, WangH09}. 
These halos with close massive companions are, therefore, expected to have  
an early assembly time because of the stripping, and can 
contribute significantly to the secondary bias 
\citep{WangH09, LiR13, Mansfield20, Tucci20}. 

In addition to the truncation and stripping processes, 
\cite{WangH11} found that dense environments can also enhance 
mass accretion by halos \citep[see also][]{Fakhouri09}. 
This process is expected to yield a trend of the halo 
bias with the assembly time that is different from the measured
secondary bias. These authors suggested that the halo-mass 
dependence of the secondary bias is partly caused by the 
competition between two categories of processes, 
the truncation by large-scale tidal field and 
the availability of material to be accreted by halos   
\citep[see also][]{Chen20}.
However, exactly how these processes contribute to the 
secondary bias is still unclear.

Suggestions have been made that the secondary bias in 
the halo concentration may share a common origin with the 
secondary bias in the assembly time \citep{Han19, Chen20}, 
although the two properties exhibit different trends as discussed above. 
Several studies \citep{Salcedo2018, Han19, Johnson19, Tucci20} pointed 
out that the secondary bias for the spin may have a 
different origin than both the assembly time and concentration. 
The correlation and alignment of the spin amplitude and direction
with the local tidal field suggest that the tidal torques may play 
a key role in establishing the dependence of the halo bias on the spin
parameter \citep{Hahn07, Shi15, Chen16, WangP2018}. 
Unfortunately, it is still unclear how the secondary bias for
the concentration and spin is established and whether 
it is related to the halo assembly or it is produced by completely 
different processes. 

Attempts have also been made to understand the origin of the 
secondary bias from the initial conditions 
\citep[e.g.][]{WangH07, Zentner07, Sandvik07, Dalal08, Desjacques08, Musso12, shi18}. 
For example, \cite{WangH07} found that low-mass, older halos tend 
to be associated with perturbations of higher mass 
in the initial density field (the initial mass) 
that are expected to collapse into halos according to the 
spherical collapse model. They further found that the bias relation 
obtained from the initial masses in $N$-body simulations 
actually matches the secondary bias in the assembly time. 
\cite{Dalal08} used the properties of density peaks 
in the initial conditions to infer the halo assembly time 
and concentration, and found that the general trends in the 
secondary bias of these two halo properties can be reproduced 
in their model. For massive halos, they suggested that the secondary 
bias reflects the statistics of the random Gaussian 
field \citep[see also][]{Zentner07}, while for low-mass
halos, they reached conclusions that are similar to those 
in earlier studies \citep[e.g.][]{WangH07,WangH09}. 
More recently using the excursion set approach, \cite{shi18}
suggested that the secondary bias possibly reflects the
correlation of densities at different scales when the
density at the halo-mass scale is fixed. 

All these results provide valuable insight into the origin of 
the secondary bias. However, the correlations between  
peak properties and halo properties are ambiguous, and the 
excursion set approach is not able to model the effects of 
the tidal truncation. 

Thus, the details of the secondary bias, 
in particular its mass dependence, remain unresolved.
It is still unclear whether or not 
the secondary bias for low- and high-mass halos have the same 
origin. It is also unclear why different halo properties exhibit 
different trends in their secondary bias, although they are correlated.  
Furthermore, since the secondary bias may already be present 
in the initial conditions, it is important to understand 
how it is connected to the secondary bias observed for the halo 
population in the evolved density field, and whether 
the initial condition or later evolution plays the more  
important role in determining the secondary bias. 

In this paper, we use both the 
evolved density field (at $z=0$) and the linear density 
field to study the secondary bias of three halo 
properties: assembly time, concentration and spin.
We use various correlation analyses to disentangle different 
effects. The paper is organized as 
follows. In Section \ref{sec_sim}, we describe the 
simulations, dark matter halo samples, the merger tree construction, 
and the quantities we use for our analyses. 
In section \ref{sec_hbel}, we study the correlation of halo
properties with the evolved and linear density fields at a
typical large scale where halo bias can be measured.
In Section \ref{sec_LEe}, we study the correlation of halo properties
with linear densities at various scales and present our findings 
of two types of correlation that can affect the secondary bias of halos. 
In Section \ref{sec_orgsb}, we use the two types of 
correlation to interpret the secondary bias for different halo 
properties and its dependence on halo mass.
Finally, we summarize and discuss our results in 
Section \ref{sec_sum}.

\section{Simulations and Dark Matter Halos}
\label{sec_sim}

\subsection{Simulations, Halos and Merger trees}\label{sec_hmt}

Two simulations with different mass resolutions are used in this paper. 
The higher resolution one is the ELUCID simulation carried out by 
\cite{WangH16} using the L-GADGET code, a memory-optimized version of 
GADGET-2 \citep{Springel05}. This simulation has $3072^{3}$ dark matter 
particles, each with a mass of $3.08\times10^{8}\msun$, 
in a periodic cubic box of 500 comoving $\mpc$ on a side. 
The other simulation has $2048^{3}$ particles in a cubic box of $1 h^{-1}$Gpc on a side,
with particle mass of $8.3\times10^{9}\msun$. 
This simulation is referred to as S1k in the following.
The initial conditions of the two simulations are generated at 
redshift of 100 by using the Zel'dovich approximation \citep{ZelDovich70}.
The cosmology parameters used in the simulations are both based on 
WMAP5 \citep{Dunkley09}: $\Omega_{\Lambda,0}$ = 0.742, $\Omega_{m,0}$ = 0.258, 
$\Omega_{b,0}$ = 0.044, h = $\rm H_0/100\kms\Mpc$ = 0.72, 
$\sigma_{8}$ = 0.80, and $n_{s}$ = 0.96. 
The characteristic collapse mass, $M_*$, defined as the characteristic mass scale 
at which the RMS of the linear density field is equal to 1.686 at the present time. For the
present simulations, $M_*$ = $10^{12.5}\msun$ at $z=0$. 
Outputs of the two simulations are made at 100 snapshots, 
from $z = 18.4$ to $z = 0$ equally spaced in the logarithm 
of the expansion factor.

Dark matter halos are identified 
using a friends-of-friends (FOF) group-finder with a linking 
length $b$ = 0.2 \citep{Davis85}. We use the SUBFIND 
algorithm \citep{Springel01} to identify gravitationally 
bound substructures (subhalos) within each FOF halo. 
This in turn makes it possible to build up halo merger trees 
to represent the detailed assembly histories of individual halos. 
Each member particle of a subhalo is assigned a weight 
that decreases with the absolute value of its binding energy. 
For a subhalo `A' in a snapshot, its descendant is identified as 
the subhalo that is in the subsequent snapshot and contains the 
largest weighted number of particles belonging to `A', and 
`A' is considered as the progenitor of its descendant.
In each FOF halo, the most massive subhalo is referred 
to as the main halo, and the branch that traces the main 
progenitors of the main halo back in time is referred to as 
the main trunk of the merging tree.

Using halo merger trees, we can also identify splashback
halos \citep[e.g.][]{Ludlow09, WangH09}, which are the main 
halos at $z=0$ but have ever been accreted by other 
massive halos in the past. As shown in Section \ref{sec_hbel}, 
splashback halos mainly affect the secondary bias for small halos.
These halos are expected to have experienced 
strong non-linear processes, in contrast to other normal halos.  
Since they are only a small fraction of the total halo population, 
we do not consider them in most of our analyses. 

We select three representative halo samples at $z=0$ to show 
our main results. 
 The first two, selected from ELUCID, contain 392,797 halos with 
$11.4\leq\log M_{\rm h}/\msun\leq 11.6$ 
(representing low-mass halos) and 49,968 halos with 
$12.4\leq \log M_{\rm h}/\msun\leq 12.6$ (representing $M_*$ halos), 
respectively. The other one contains 9,085 halos with 
$13.9\leq \log M_{\rm h}/\msun\leq 14.1$ 
(representing massive halos) selected from the S1k simulation. 
The results for other halo mass bins are presented when necessary.
Note that we only show results for halos that each contain more 
than 800 dark matter particles, corresponding to 
$\log(M_{\rm h}/\msun)>11.4$ in ELUCID and $\log(M_{\rm h}/\msun)>12.9$ in S1k.
As shown below, our results using ELUCID and S1k, which have  
very different mass resolutions, are similar, suggesting that
mass resolution does not affect our conclusions significantly.

\subsection{Halo properties and overdensities}\label{sec_pq}

\begin{figure*}
    \centering
    \includegraphics[scale=0.5]{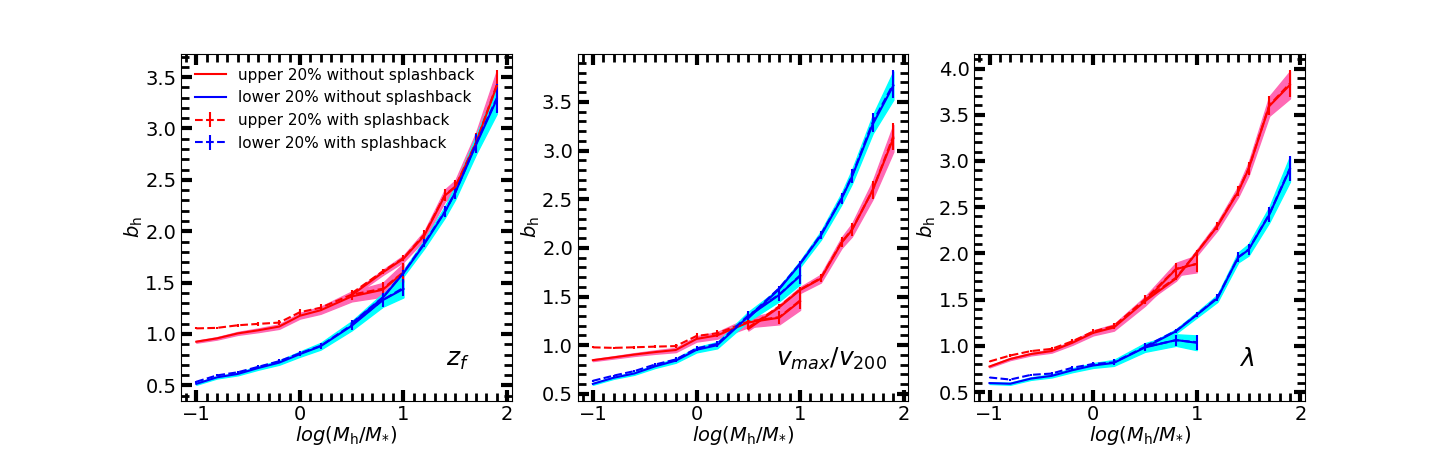}
	\caption{Halo bias factor as a function of halo mass and other halo properties. In each panel 
	the red (blue) lines show the results for halos in the upper (lower) 20\% tails of the distribution of the halo property indicated in the panel. The dashed line shows the result for all halos, while the solid line shows the result excluding splashback halos.  Results are shown for ELUCID and S1k halos in the mass ranges of $\log(M_{\rm h}/M_{*})<1$ and $\log(M_{\rm h}/M_{*})>0.5$, respectively. Error bars are the standard deviation calculated using 1000 bootstrap samples. }
	\label{fig_bias}
\end{figure*}

\begin{figure*}
    \centering
    \includegraphics[scale=0.5]{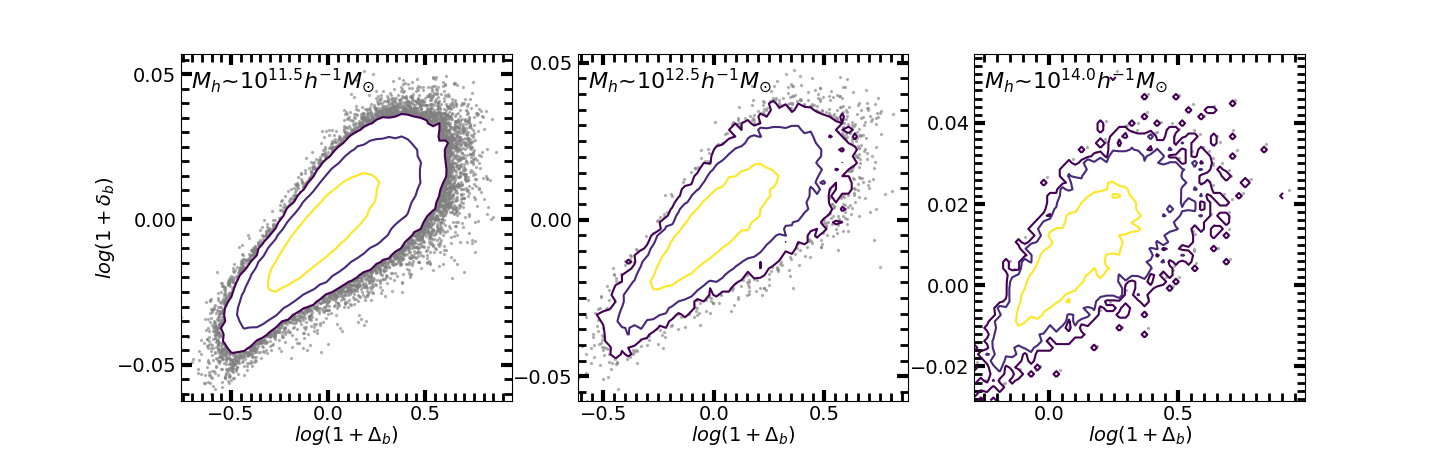}
	\caption{ The correlations between the overdensities measured at $z=18.4$ ($\delta_{\rm b}$) and at $z=0$ ($\Delta_{\rm b}$) for three representative halo samples, as indicated in the panels. Both overdensities are measured within a large comoving radius range, $R_{\rm b}=[10-15]\mpc$. The three  contour lines in each panel enclose 67\%, 95\% and 99\% of halos, respectively.}
	\label{fig_dbdib}
\end{figure*}

\begin{figure} 
    \centering
    \includegraphics[scale=0.4]{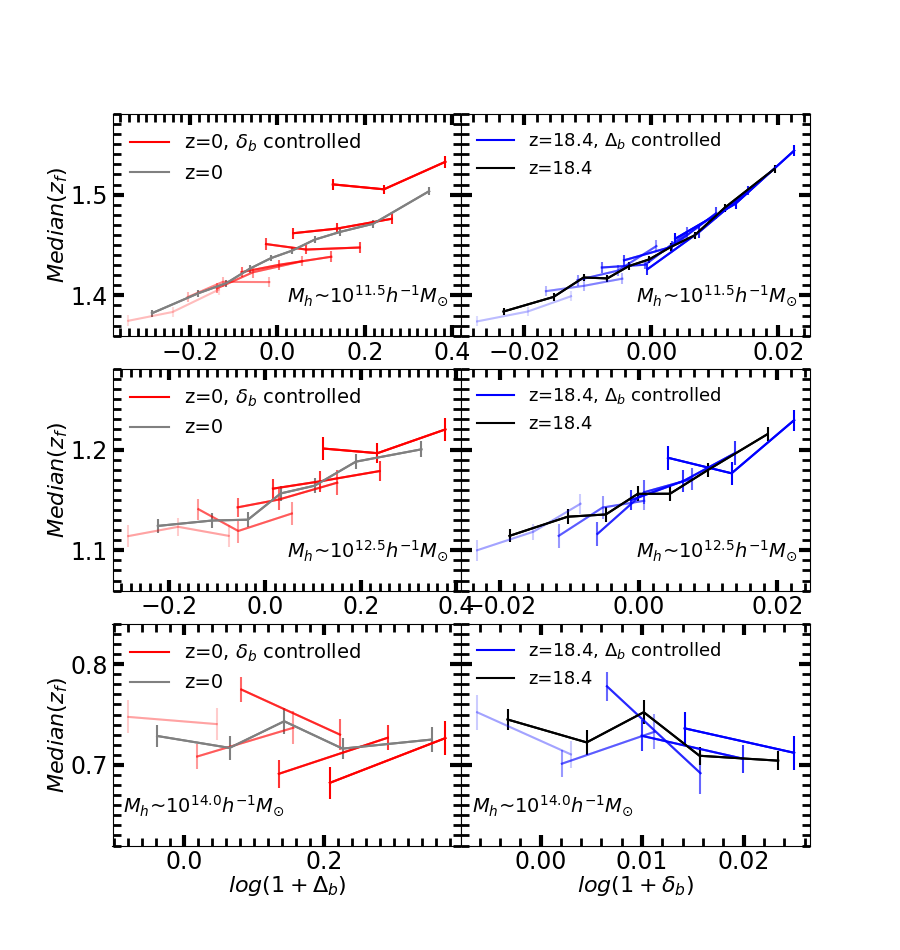}
	\caption{ The median $\zf$ as a function of $\Delta_{\rm b}$ (left panels)
	and $\delta_{\rm b}$ (right panels) for three representative halo samples, as indicated in the panels. 
	The red (blue) lines in left (right) panels show the results with $\delta_{\rm b}$ 
	($\Delta_{\rm b}$) controlled (see the text for details). The error bars show the standard deviation calculated 
	using 1000 bootstrap samples.}
	\label{fig_zfdib}
\end{figure}

In this paper, we focus on the secondary bias of halo distribution 
using three halo properties: the assembly time, $\vr$ and the spin. 
Here we list the definitions of these three properties together 
with the halo mass:
\begin{itemize}
\item Halo mass $M_{\rm h}$: the mass contained in the spherical region of radius
$r_{200}$, centered on the most bound 
particle of the main halo, and within which the mean mass density is equal to 200 times the critical density.
\item Halo assembly time $z_{\rm f}$: the redshift at which $M_{\rm h}$ reaches half of its 
final mass at $z=0$. It is determined by tracing the main trunk of the merger tree of the halo in question.
\item $v_{\rm max}/v_{\rm 200}$: the ratio of the peak value of its circular velocity profile to the virial 
velocity. Here virial velocity is defined as the circular velocity at $r_{200}$. 
This parameter is often used to characterize the concentration of a halo \citep[e.g.][]{Gao07}.
\item Halo spin $\lambda$: defined as $\lambda=|{\bf J}|/(\sqrt{2}M_{\rm h}v_{200}r_{200})$, 
where $J$ is the angular momentum measured by using particles within a sphere of $r_{200}$.
\end{itemize}

 As mentioned in the introduction, for a given halo mass, the halo bias depends   
significantly on the other three halo properties. To avoid ambiguity,   
we refer to the dependencies as the $\zf$ bias, the $\vr$ bias and the $\lambda$ bias, respectively.
To understand their origin, we use over-densities measured on different scales
from both the evolved and linear density fields. We thus need to know the positions 
of the proto-halos that correspond to the halos identified at $z=0$.
For each FOF halo at $z=0$,  the position of the proto-halo is defined  
as the average position of all particles that end up in the $z=0$ halo.
The following is the list of the over-densities used in our analyses:
\begin{itemize}
\item $\Delta_{\rm b}$, the overdensity measured within a large comoving radius range, 
$R_{\rm b}=[10-15]\mpc$, centered on each halo {\it at $z=0$}. It is used to infer the halo bias 
factor at $z=0$.
\item $\delta(R)$, the linear overdensity at $z=18.4$ at a series of 
comoving radius $R$, centered on the position of a proto-halo. 
\item $\delta_{\rm b}$,  the linear overdensity at the comoving radius 
$R_{\rm b}$ at $z=18.4$, centered on the position of a proto-halo. Note that 
$\delta_{\rm b}$ is exactly $\delta(R)$ when $R=R_{\rm b}$.
\item $\delta_{\rm L}$, the linear overdensity at $z=18.4$ within the halo 
Lagrangian radius, $R_{\rm L}$, centered on the position of its 
proto-halo. $R_{\rm L}\equiv (M_{\rm h}/(4\pi/3\bar\rho))^{1/3}$, where $M_{\rm h}$ 
is the halo mass of the corresponding halo at $z=0$ and $\bar\rho$ is the mean 
comoving density of the universe. Note that
$\delta_{\rm L}$ is exactly $\delta(R)$ when $R$ is chosen in the 
range of $[0, R_{\rm L}]$.
 \item $\delta_{\rm e}$, the linear overdensity at $z=18.4$ within $[1,1.2]R_{\rm L}$ 
for a proto-halo. It is exactly $\delta(R)$ when $R=[R_{\rm L}, 1.2 R_{\rm L}]$.
\end{itemize}

For clarity, we use $\delta$ to denote the overdensity measured at $z=18.4$ and $\Delta$ to denote the present-day overdensity.
As shown in \cite{Han19}, the bias factor at the scale 
from 5 to $20\mpc$ is well consistent with the linear theory. 
We thus adopt $R_{\rm b}=[10,15]\mpc$ to measure $\Delta_{\rm b}$.
For any given halo sample, the halo bias is calculated as 
\begin{equation}
   b_{\rm h}=\frac{\left<\Delta_{\rm b}\right>}{\bar\Delta_{\rm b, p}}\,
\end{equation}
where $\left<\cdot\cdot\cdot \right>$ denotes the average over 
all halos in the sample, and $\bar\Delta_{\rm b, p}$ is the mean 
overdensity measured at $R_{\rm b}$ centered on all particles
in the density field  at $z=0$.

The linear densities used in this paper are measured 
from the snapshot at $z=18.4$ rather than the initial 
condition at $z=100$. At $z=100$, the true overdensity is 
very small and the shot noise of particles is important, 
which can lead to systematic bias in the estimate of the 
true linear density fluctuation. 
On the other hand, the overdensities obtained at $z=18.4$ 
are in good agreement with the prediction of the linear 
perturbation theory (see Appendix).  Finally, we note that the Lagrangian radii, 
$R_{\rm L}$, for the three representative halo samples 
are about 1.02, 2.19 and 6.94 $\mpc$, respectively.

\section{Halo secondary bias in evolved and linear density fields}\label{sec_hbel}

\begin{figure}
    \centering
    \includegraphics[scale=0.4]{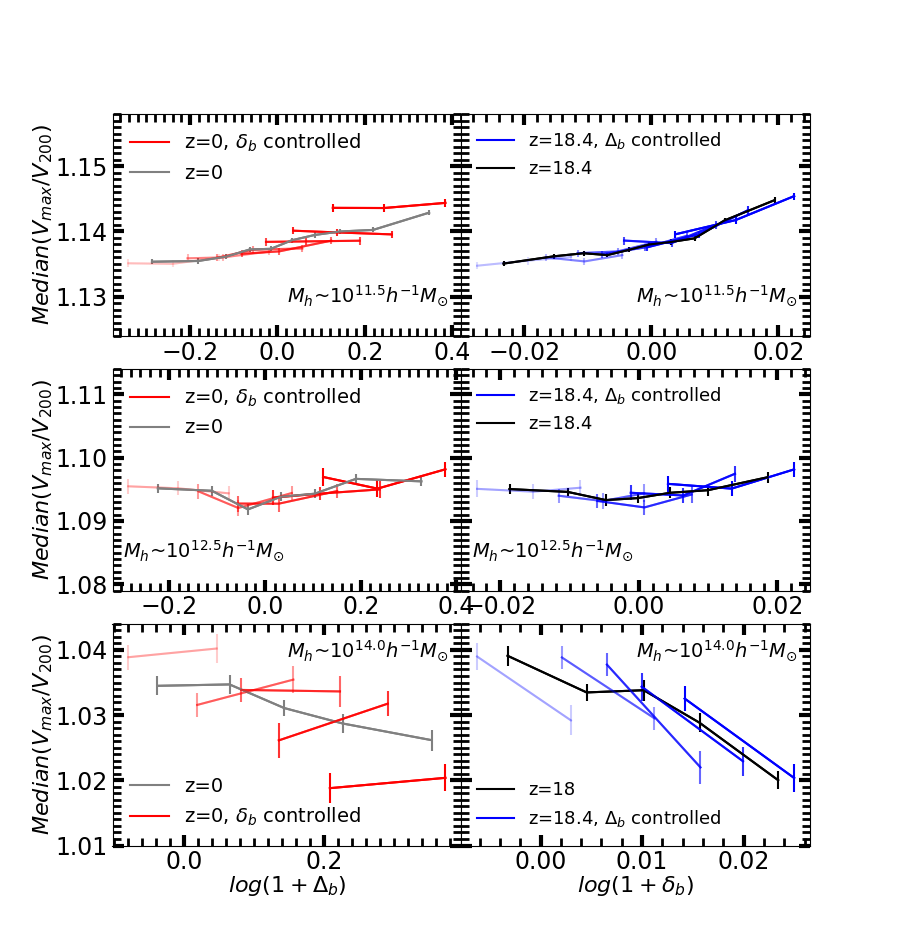}
	\caption{The same as Fig. \ref{fig_zfdib} but for $\vr$.}
	\label{fig_vrdib}
\end{figure}

Fig. \ref{fig_bias} shows the secondary bias for 
the three halo properties,
$z_{\rm f}$ (left), $\vr$ (middle) and 
$\lambda$ (right). In a given narrow mass range (0.2 dex), 
we select two sub-samples that consist of halos in the lower and upper 20 
percent tails of the distribution of the halo property in question, 
respectively. We calculate the halo bias ($b_{\rm h}$) for each of 
the sub-samples and show it as a function of halo mass. The dashed and solid lines show the 
results including and excluding splashback halos, respectively. 
 For the ELUCID simulation, we present results for halos with
$-1\leq\log(M_{\rm h}/M_{*})\leq1$, while for the S1k simulation,
results are shown for $\log(M_{\rm h}/M_{*})\geq0.5$.
Although the two simulations have very different 
mass resolutions, their results agree with each other 
well in the overlapping mass range, suggesting that
our results are not significantly affected by numerical
resolutions.

Fig. \ref{fig_bias} shows that splashback halos 
mainly affect results for low-mass halos, 
as expected from the fact that the fraction of the 
splashback population decreases with increasing halo 
mass \citep[][]{WangH09}. The exclusion of splashback halos
decreases the bias for halos in the upper percentiles of the 
$z_{\rm f}$ and $\vr$ distributions,  
while the halo bias for the lower percentiles is not 
affected significantly. This is expected. At low $\zf$, halo bias 
depends only weakly on $\zf$ \citep[see e.g.][]{Gao05}, while  
splashback halos on average have higher $\zf$ \citep[e.g.][]{WangH09}.  
Different from $\zf$ and $\vr$, the $\lambda$ bias for the two subsamples are
both significantly affected by the splashback halos.
Splashback halos are expected to have experienced strong non-linear evolution, 
and consequently behave very differently from other halos in 
their relations to the linear density field.
To reduce uncertainties caused by splashback halos, we exclude them in our analyses.

All the three parameters show strong secondary bias, 
but with very different halo-mass dependence
\citep[see also e.g.][]{Faltenbacher2010, Salcedo2018}. 
Older halos are usually more strongly clustered than younger
ones of the same mass. However, the $z_{\rm f}$ bias 
becomes weaker as halo mass increases and is almost absent 
at $\log(M_{\rm h}/M_*)>1$.
The $\vr$ bias is significant over the whole 
mass range covered. More interestingly, it changes sign around 
$M_h\sim M_*$, above which less concentrated halos are actually more strongly 
biased. The $\lambda$ bias is also strong in the whole mass range. 
Different from the other two parameters, its strength increases 
with the halo mass. 

It is known that $\vr$ increases with $z_{\rm f}$
at given halo mass \citep[e.g.][]{Gao04, Han19}, 
and so the $\zf$ bias and $\vr$ bias may have a 
similar origin for low mass halos, 
as suggested by previous studies \citep[e.g.][]{Chen20}.
However, for massive halos, the two secondary biases behave 
differently from the expectation of the correlation between 
the two parameters. Moreover, it is also known that older 
halos tend to have smaller spin over the whole mass range
\citep[e.g.][]{Hahn07,WangH11}, indicating that the $z_{\rm f}$ bias and 
$\lambda$ bias must be caused by different processes.

\begin{figure}
    \centering
    \includegraphics[scale=0.4]{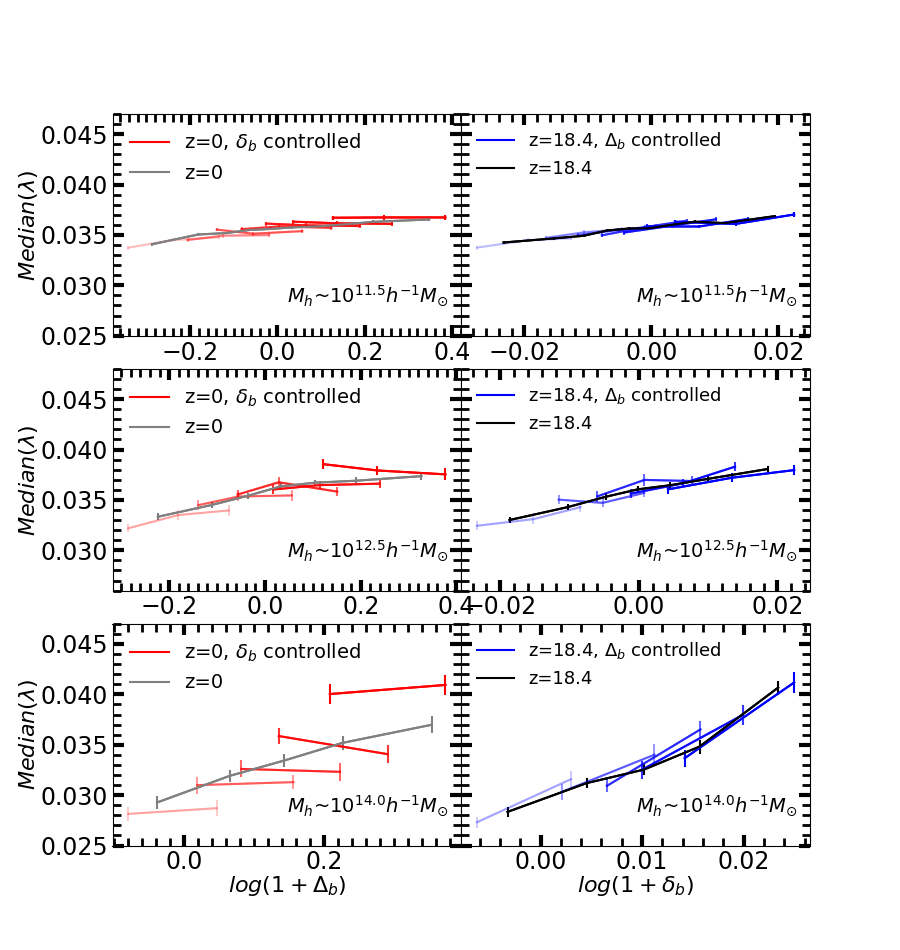}
	\caption{Similar to Fig. \ref{fig_zfdib} but for spin, $\lambda$.}
	\label{fig_spdib}
\end{figure}

To understand the origin of the secondary bias and the 
complex mass dependence shown above, we investigate the correlations 
of the halo properties with the linear density field. 
Here we first focus on $\delta_{\rm b}$, 
which is estimated at the same comoving scale as $\Delta_{\rm b}$. 
 As shown in Fig. \ref{fig_dbdib}, $\delta_{\rm b}$ and
$\Delta_{\rm b}$ are strongly correlated, albeit with 
considerable variance. It is thus interesting to investigate whether 
their correlations with halo properties are similar.
We show the median of halo properties as functions of 
$\Delta_{\rm b}$ and $\delta_{\rm b}$ for the three representative halo samples 
in Figs. \ref{fig_zfdib}, \ref{fig_vrdib} and \ref{fig_spdib}.
To see which density indicator dominates the correlations, 
we split each of the representative halo samples 
into several equal-sized sub-samples according to $\Delta_{\rm b}$
or $\delta_{\rm b}$. We then show halo properties of individual subsamples of fixed 
$\delta_{\rm b}$ as a function of $\Delta_{\rm b}$
in the left panels (red lines), and those of fixed $\Delta_{\rm b}$
as a function of $\delta_{\rm b}$ in the right panels (blue lines). 
The numbers of the equal-sized subsamples are 7, 5 and 5 for 
the low-mass, $M_*$ and cluster-sized halos, respectively.

\begin{figure*}
    \centering
    \includegraphics[scale=0.5]{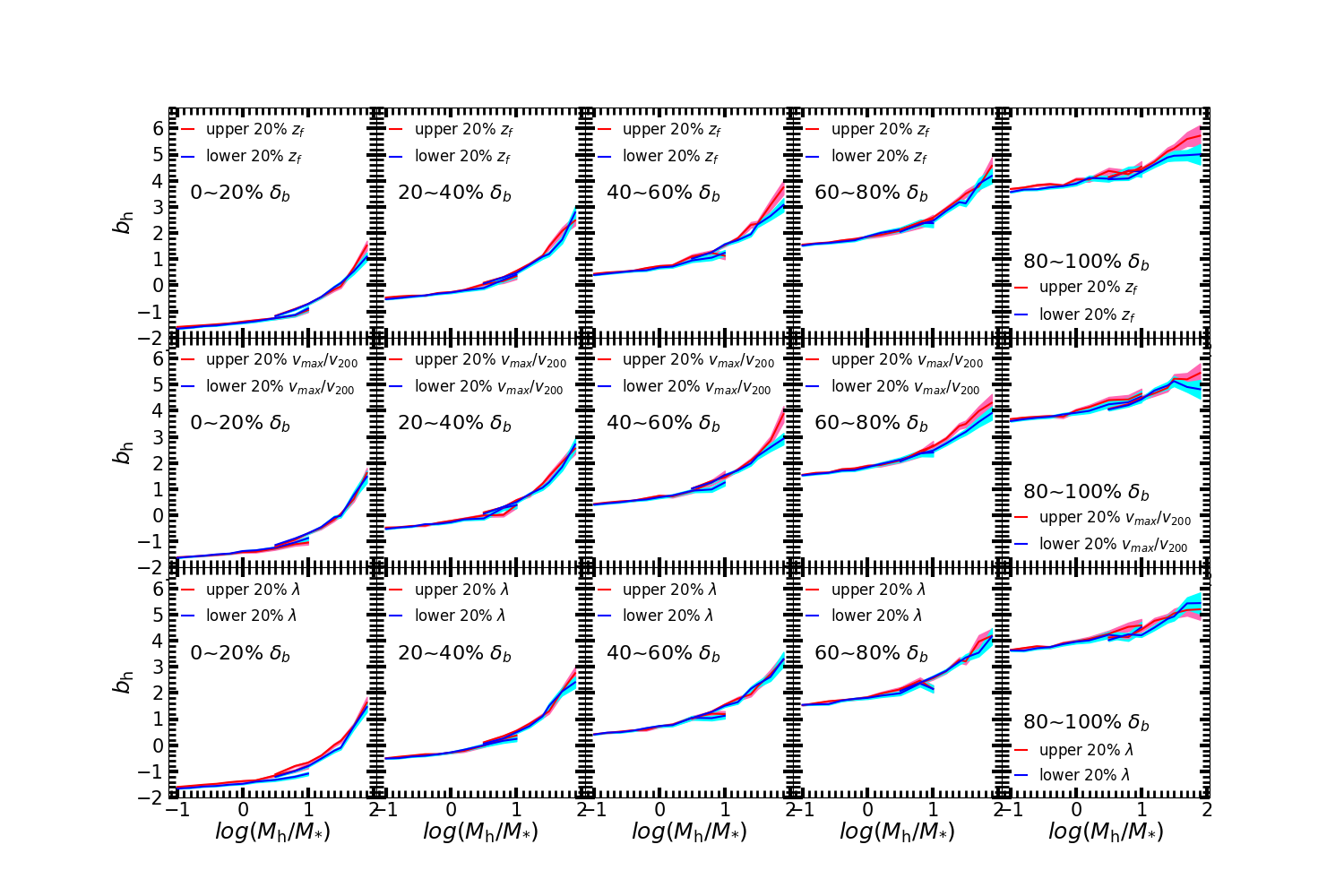}
	\caption{Halo bias factor as a function of $M_{\rm h}$, $\delta_{\rm b}$ and halo properties 
	(upper panels: $\zf$, middle panels: $\vr$, and lower panels: $\lambda$).
	 For a given halo mass bin, we divide halos into 5 subsamples of equal size 
	 according to $\delta_{\rm b}$, as indicated in different columns.
	The red (blue) lines show the results for halos in the upper (lower) 20\% of the distribution 
	of a given halo property with $\delta_{\rm b}$ controlled. The symbols, color-code and shaded 
	region are the same as in Fig. \ref{fig_bias}.}
	\label{fig_bias_cdib}
\end{figure*}

For the halo assembly time, one can see clear dependence
on both $\Delta_{\rm b}$ and $\delta_{\rm b}$. 
The overall correlation strengths with the two densities are similar.
The dependence weakens with increasing halo mass and 
disappears for cluster-sized halos, consistent with the results 
shown in Fig. \ref{fig_bias}. 
As one can see, for the two low mass bins, 
the dependence on $\Delta_{\rm b}$ becomes very weak
when $\delta_{\rm b}$ is controlled. 
On the other hand, when $\Delta_{\rm b}$ is controlled, 
the correlation of $\zf$ with $\delta_{\rm b}$
for individual sub-samples follows the overall trend closely.
For the most massive halos, no clear trend can be seen. 

The results for $\vr$ are also consistent with those
shown in Fig. \ref{fig_bias}. One sees that the correlation
shows opposite trends for low-mass and the most massive 
halos. The overall trends with $\Delta_{\rm b}$ and $\delta_{\rm b}$ 
are similar. However, as shown in Fig. \ref{fig_vrdib}, 
for both low and high mass halos, the
dependence on $\Delta_{\rm b}$ is absent when 
$\delta_{\rm b}$ is controlled, but a clear 
correlation (positive for low-mass halos and negative for 
high-mass halos) is clearly seen with $\delta_{\rm b}$ 
for fixed $\Delta_{\rm b}$. For halos with $M_h\sim M_*$,
on the other hand, the trends of $\vr$ with $\delta_{\rm b}$ 
and $\Delta_{\rm b}$ are both rather weak.

Finally, the overall correlation of $\lambda$ with the two density 
parameters becomes stronger as the halo mass increases, 
consistent with the results shown in Fig. \ref{fig_bias}. 
If samples are not controlled, the general trend in  
the $\lambda$-$\Delta_{\rm b}$ correlation is similar to that 
in the $\lambda$-$\delta_{\rm b}$ correlation. However,  
the results of the controlled samples clearly show that the linear density, 
$\delta_{\rm b}$, is the driving factor of the $\lambda$-$\Delta_{\rm b}$ 
correlation for all the three halo samples.

The analyses presented above demonstrate clearly that 
the correlations of halo properties with $\Delta_{\rm b}$ 
are driven by the correlations with $\delta_{\rm b}$ in all 
cases where significant secondary bias is detected, through the 
correlation between $\Delta_{\rm b}$ and $\delta_{\rm b}$.
To demonstrate this in a more intuitive way, we divide halos in each mass bin 
into 5 subsamples of equal size according to $\delta_{\rm b}$. We then select halos 
in the upper and lower 20\% of the distribution of a given  halo property in each 
subsample and show the corresponding halo bias as a function of halo mass in 
Fig. \ref{fig_bias_cdib}. As one can see, when $\delta_{\rm b}$ is controlled, the secondary bias 
for all the three halo properties disappears over the entire halo mass range 
probed, consistent with the results shown in Fig. \ref{fig_zfdib}, \ref{fig_vrdib} and \ref{fig_spdib}. 
This demonstrates again that the secondary bias is more closely tied to the linear 
density field than to the halo properties.
This also suggest that the origin of the secondary 
bias may be approached by investigating the correlations of 
halo properties with the linear density field, as we will do below.

\section{Correlations of halo properties with linear densities on various scales}\label{sec_LEe}

\begin{figure*}
    \centering
    \includegraphics[scale=0.5]{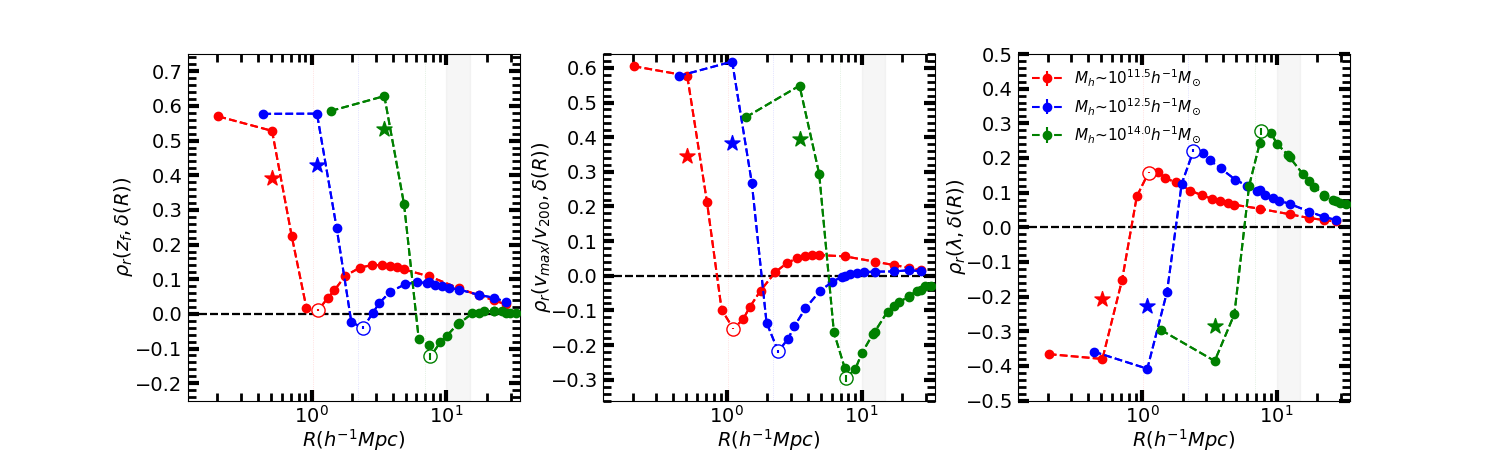}
	\caption{Pearson correlation coefficients ($\rho_{\rm r}$) between halo properties (left panel: $\zf$; middle panel: $\vr$; right panel: $\lambda$) and $\delta(R)$ as a function of $R$ for three 
	representative halo samples, as indicated. 
	The vertical error bars (usually too small to be seen) indicate the uncertainties of the coefficients, 
	calculated using 200 bootstrap samples. The results for $\delta_{\rm L}$ 
	and $\delta_{\rm e}$ are marked with stars and open circles, respectively. 
	Results on other scales are shown with solid circles. 
	The vertical dotted lines indicate $R_{\rm L}$ of the halos in question.
	The vertical shaded region indicates the range of $R_{\rm b}$ used to estimate the halo bias.}
	\label{fig_hpdcor}
\end{figure*}

\begin{figure*}
    \centering
    \includegraphics[scale=0.5]{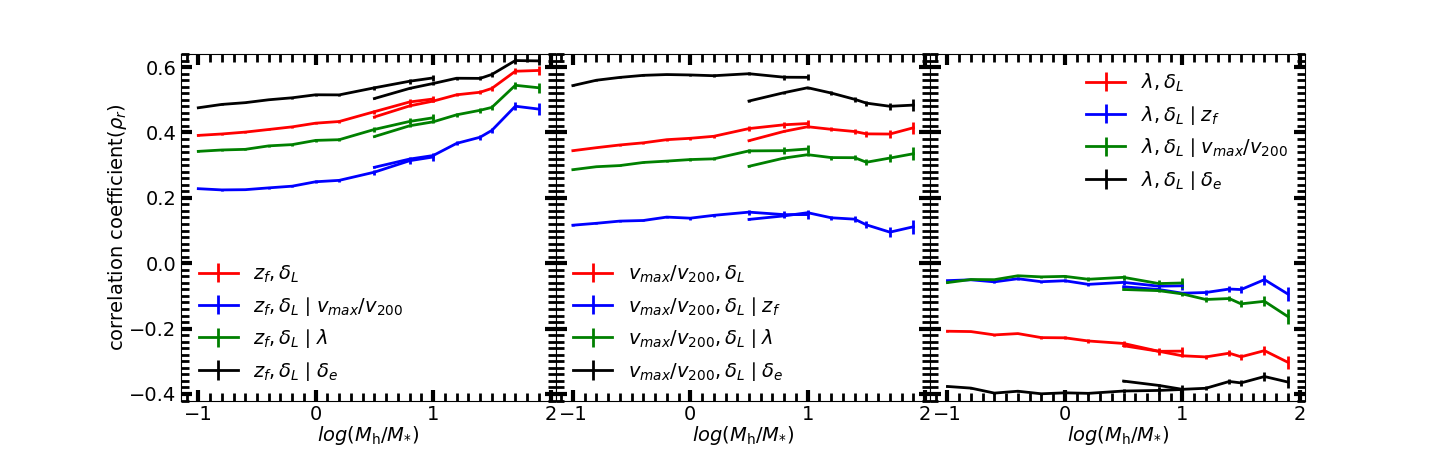}
	\caption{Pearson coefficients for the (partial) correlations of $\delta_{\rm L}$ with halo properties, $\zf$ (left panel), $\vr$ (middle panel) and $\lambda$ (right panel),  as a function of halo mass. As indicated in each panel, the red lines show the results of the correlations between $\delta_{\rm L}$ and halo properties, the blue and green lines show the results with another property controlled, and the black lines show the results with $\delta_{\rm e}$ controlled. The error bar shows the standard deviation calculated from 200 bootstrap samples.}\label{fig_hpdiL}
\end{figure*}

The results presented above show that halo properties are 
correlated with the linear over-density at a large 
comoving scale, $R_{\rm b}$. \cite{WangH07} found 
that the halo assembly time is also correlated with the linear 
over-density measured within the Lagrangian radius, 
$\delta_{\rm L}$. These two results suggest that 
halo properties are correlated with the linear density field 
on various scales. It is thus interesting to first investigate 
in more detail how the correlation varies with the scale.
A convenient way to characterise the correlation strength of  
two variables, $x$ and $y$, is to use their Pearson correlation 
coefficient, defined as
\begin{equation}\label{eq_rhor}
    \rho_{\rm r}(x,y)=\frac{\left<(x-\bar x)(y-\bar y)\right>}{\sigma_x\sigma_y}\,
\end{equation}
where $\bar x$ ($\bar y$) and $\sigma_x$ ($\sigma_y$) are the mean and 
standard deviation of the variable $x$ ($y$), respectively. 
As discussed in \cite{Han19}, the Pearson coefficient is a powerful 
tool for correlation analysis. If the two variables are roughly 
linearly correlated, the coefficient measures the steepness of 
the correlation between the corresponding normalized variables. 

The Pearson coefficients, $\rho_{\rm r}$, 
for the correlations of halo properties with $\delta(R)$ 
are shown in Fig. \ref{fig_hpdcor}.
Let us first look at the coefficients at the scale $R=R_{\rm b}$
(indicated by the vertical bands), 
i.e. the correlations of halo properties with $\delta(R_{\rm b})=\delta_{\rm b}$,
and compare them with the results shown in the last section. 
The absolute values of the correlation coefficients of the three halo 
properties with $\delta_{\rm b}$ range from 0 to 0.2,
indicating that the secondary bias is in general a weak effect. 
The results show that $\delta_{\rm b}$ is positively correlated 
with $\zf$ for the two low-mass samples, and that the coefficient 
is close to zero for the most massive halos. The coefficient for $\vr$ 
changes its sign around $M_*$, being positive and negative for 
halos of lower and higher masses, respectively. 
The correlation coefficient for $\lambda$ is always positive and 
increases with halo mass. All of these are in good agreement with the 
results shown in the last section.

\begin{figure*}
    \centering
    \includegraphics[scale=0.5]{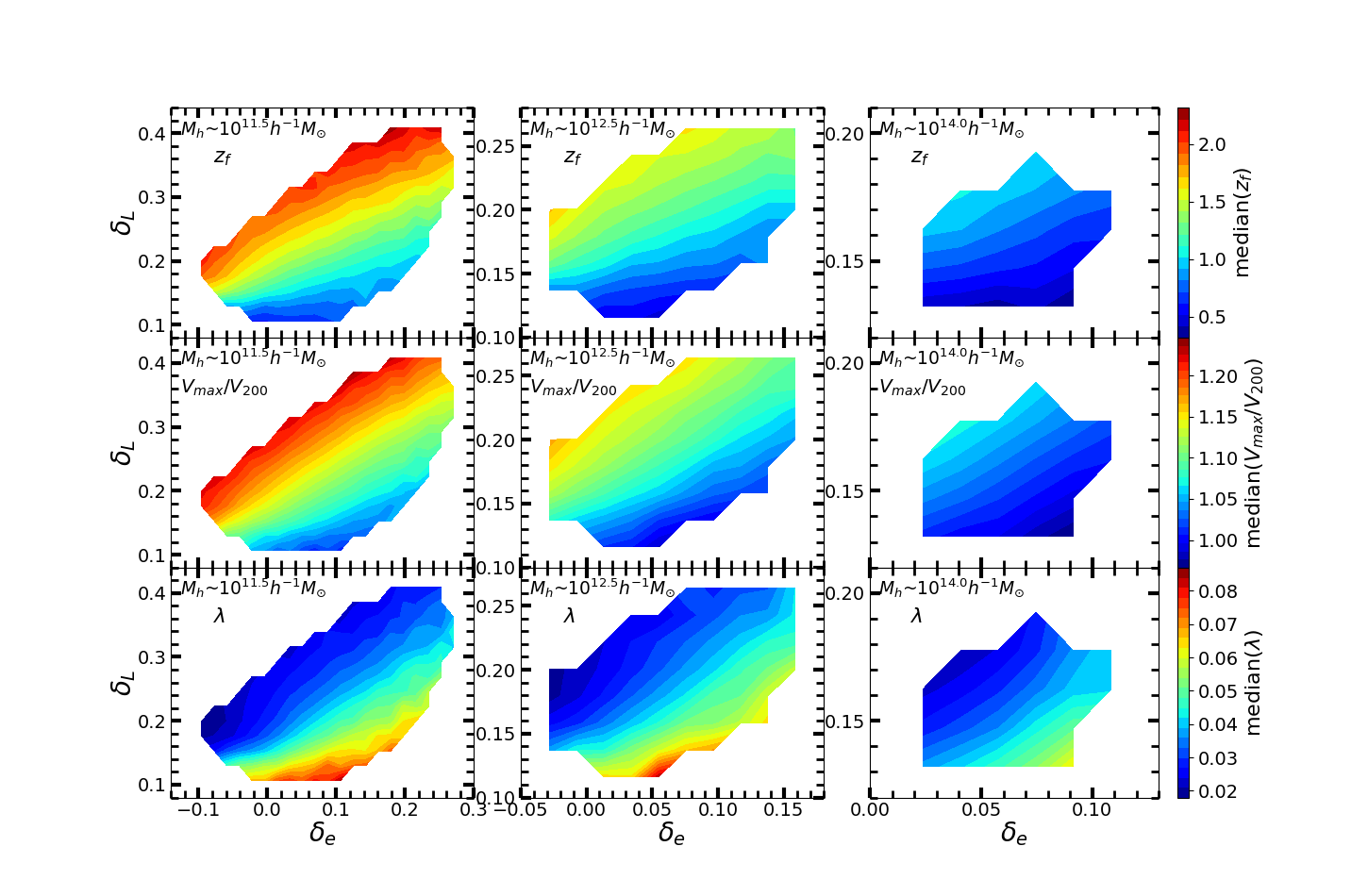}
	\caption{Contours show median $z_{\rm f}$ (upper panels), $\vr$ (middle panels), $\lambda$ (lower panels) as a function of  $\delta_{\rm e}$ and $\delta_{\rm L}$ for three representative halo samples as indicated in the panels. Only grids containing more than 100 halos are presented.}
	\label{fig_hp2d}
\end{figure*}

The scale-dependence of these correlations is complex, 
although there are some interesting features. 
In the inner region of proto-halos, i.e. 
$R\leq 0.6R_{\rm L}$, the correlations are almost independent 
of the scale. As $R$ increases, we see a rapid change in 
the correlation strength and even a change in the sign of the 
correlations around the Lagrangian radius, $R_{\rm L}$.  
At $R=[1,1.2]R_{\rm L}$, the coefficients 
reach a (local) minimum for $\zf$ and $\vr$, and a maximum 
for $\lambda$. At $R>R_{\rm L}$, the correlations 
show complex dependence on halo mass, which is different 
from the dependence on the overdensities within proto-halos.
For example, for low-mass and $M_*$ samples, the 
coefficient for $\zf$ first increases with $R$ and then decreases gradually 
to zero at large scales. For cluster-sized halos, 
on the other hand, the scale dependence appears monotonous.

These results suggest that the overdensities inside and 
outside the proto-halos may affect halo properties in 
different ways. In the following, we therefore consider the dependence 
on these two types of ovserdensities separately.
Specifically, we use $\delta_{\rm L}$, the linear overdensity within the 
Lagrangian radius, to represent the inside overdensity of proto-halos, 
and use $\delta_{\rm e}$, measured at $R=[1,1.2]R_{\rm L}$, to represent 
the outside overdensity (see Section \ref{sec_pq} for the definition of the two quantities).
We will also consider overdensities measured at larger scales when necessary.

Fig. \ref{fig_hpdiL} shows the correlation 
coefficients between the three halo properties and 
$\delta_{\rm L}$ as a function of halo mass. 
The three halo properties have different 
correlation strengths with $\delta_{\rm L}$, being the 
strongest for $\zf$ ($\rho_{\rm r}=0.4 \to 0.6$) and 
the weakest for $\lambda$ ($\rho_{\rm r}=-0.2\to -0.3$). 
The strong and positive correlation with the halo assembly time
is well consistent with what was found in \cite{WangH07}. 
It is interesting that all the correlations 
show only a weak dependence on halo mass, which looks very 
different from the complex halo mass dependence of the 
secondary bias. We will come back to this question in the next section.
Since the two simulations with very different mass resolutions
give similar coefficients in the overlapping mass range,
it indicates again that numerical effects are
not important for the halos used in our analyses.

\begin{figure}
    \centering
    \includegraphics[scale=0.6]{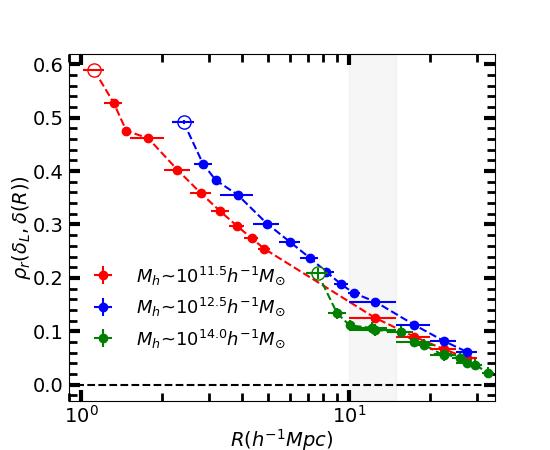}
	\caption{Pearson coefficients between $\delta_{\rm L}$ and $\delta(R)$ as a function of $R$ for three representative halo samples. The symbols,
	color-code, error bars and shaded region are the same as Fig. \ref{fig_hpdcor}.
	}\label{fig_rhocor1}
\end{figure}

In Fig. \ref{fig_hp2d}, we show the average of the halo properties in 
the $(\delta_{\rm L}$ versus $\delta_{\rm e})$ space. 
As one can see, the three halo 
properties exhibit strong dependence on both overdensities in all 
the three mass bins. Consistent with the correlation 
coefficient results shown in Fig. \ref{fig_hpdiL}, 
both $\zf$ and $\vr$ have a strong positive correlation with 
$\delta_{\rm L}$, while $\lambda$ has a negative 
correlation with $\delta_{\rm L}$.
We also see a clear {\it negative} correlation of 
both $\zf$ and $\vr$, and a clear {\it positive} correlation  
of $\lambda$, with $\delta_{\rm e}$ for given 
$\delta_{\rm L}$. These correlations are not 
always consistent with the results shown in Fig. \ref{fig_hpdcor}. 
The discrepancy is produced by the fact that $\delta_{\rm L}$ 
and $\delta_{\rm e}$ are correlated, so that the correlation 
with $\delta_{\rm e}$ depends on whether or not $\delta_{\rm L}$
is controlled. Fig. \ref{fig_rhocor1} shows the 
correlation coefficient between $\delta_{\rm L}$ and 
$\delta(R)$ as a function of $R$. Clearly, 
$\delta_{\rm L}$ is correlated with the over-density on different scales, including $\delta_{\rm e}$.
The correlation coefficient decreases gradually with $R$, and
depends only weakly on halo mass.

\begin{figure*}
    \centering
    \includegraphics[scale=0.4]{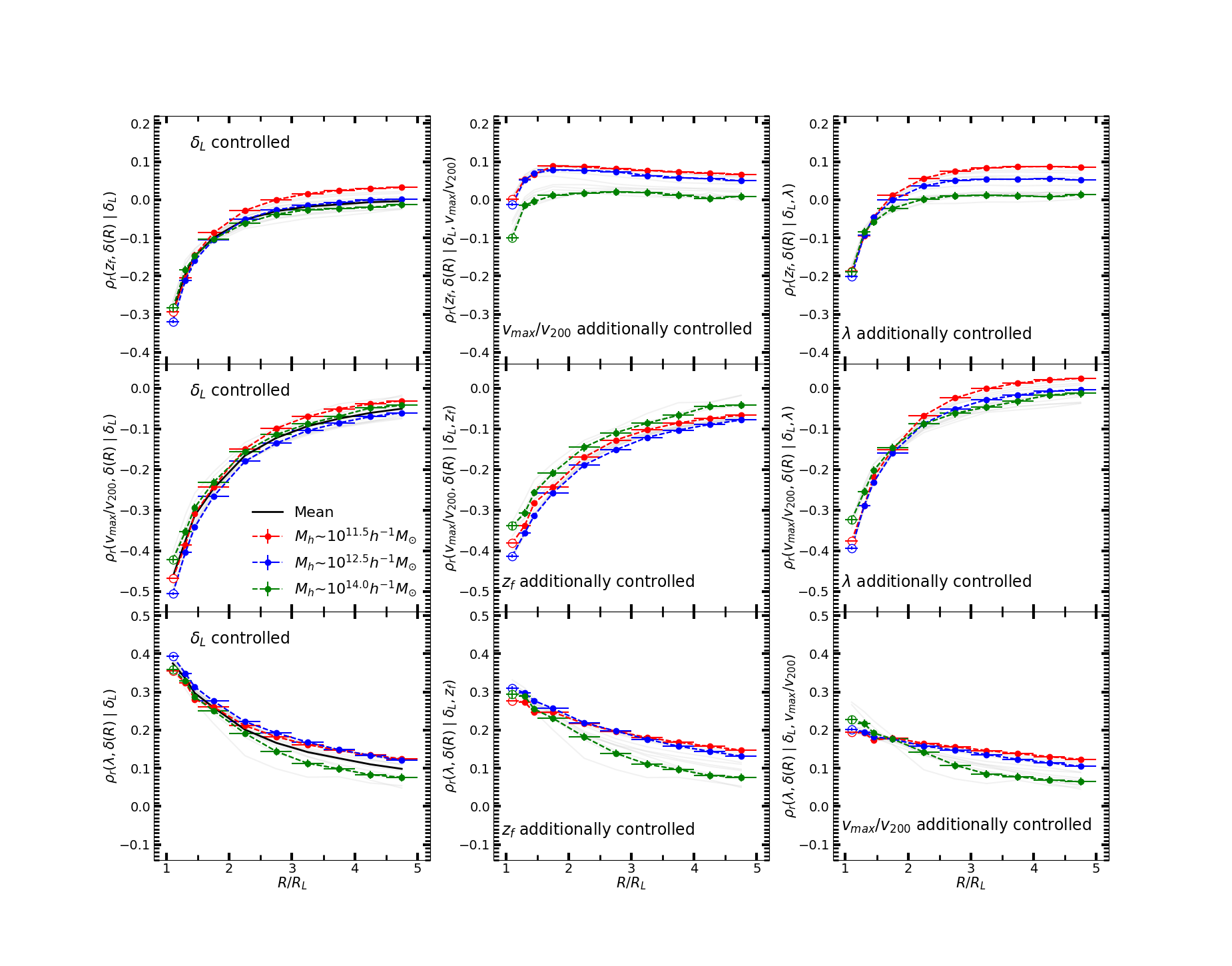}
	\caption{Partial correlation coefficients for correlations between halo properties and $\delta(R)$, with $\delta_{\rm L}$ controlled (left panel), with $\delta_{\rm L}$ and another halo property additionally
	controlled (middle and right panel), as a function of $R/R_{\rm L}$. The results in left panels are computed  using Eq. \ref{eq_pcoe} and those in the middle and right panels using Eq. \ref{eq_pcoe2}. The symbols,
	color-code and error bars are the same as Fig. \ref{fig_hpdcor}. The gray lines show the results for other halo mass bins and the black lines show the mean results of these halo mass bins.}
	\label{fig_partial}
\end{figure*}

It is thus important to disentangle these correlations, and 
we use partial correlation coefficient to do so. The partial correlation coefficient 
measures the correlation coefficient between two 
quantities, $x$ and $y$, with a third parameter, $z$, controlled, 
and is defined as 
\begin{equation}
    \rho_{\rm r}(x,y|z)=\frac{\rho_{\rm r}(x,y)-\rho_{\rm r}(x,z)\rho_{\rm r}(y,z)}{\sqrt{1-\rho^2_{\rm r}(x,z)}\sqrt{1-\rho^2_{\rm r}(y,z)}}\\.\label{eq_pcoe}
\end{equation}
\citep[see][]{John89}.
Higher-order partial coefficient can be computed
from the lower-order coefficient. For example, when two 
variables, $z$ and $w$, are controlled, the coefficient can be
written as,
\begin{equation}
    \rho_{\rm r}(x,y|z,w)=\frac{\rho_{\rm r}(x,y|z)-\rho_{\rm r}(x,w|z)\rho_{\rm r}(y,w|z)}{\sqrt{1-\rho^2_{\rm r}(x,w|z)}\sqrt{1-\rho^2_{\rm r}(y,w|z)}}\\.\label{eq_pcoe2}
\end{equation}

The correlation coefficients between the three 
halo properties and $\delta_{\rm L}$, with
$\delta_{\rm e}$ controlled, are plotted in 
Fig. \ref{fig_hpdiL}. With $\delta_{\rm e}$ controlled,
the correlations with $\delta_{\rm L}$ become significantly 
stronger. This is expected, as the two densities are positively correlated 
(Fig. \ref{fig_rhocor1}) but they have opposite correlations with 
the halo properties (Fig. \ref{fig_hp2d}). Note again that the halo mass 
dependence in these correlations are all weak.

The correlation coefficients between the halo properties 
and $\delta(R)$, with $\delta_{\rm L}$ controlled, 
are shown in Fig. \ref{fig_partial}. Here results are shown 
as a function of $R/R_{\rm L}$, instead of $R$.
Controlling $\delta_{\rm L}$ boosts 
the correlation strengths between $\delta_{\rm e}$ 
and the halo properties, and makes the coefficients  
monotonous functions of $R/R_{\rm L}$ at $R/R_{\rm L}>1$.
Among the three halo properties, $\vr$ has the 
strongest (anti-) correlation with $\delta_{\rm e}$, 
and $\zf$ the weakest. 
Moreover, the positive correlation of $\lambda$ with $\delta(R)$
extends to $R>5R_{\rm L}$. In general, the correlations 
with $\delta_{\rm e}$ and density at larger scales 
are the opposite to the correlations with $\delta_{\rm L}$,
consistent with the results shown in Fig. \ref{fig_hp2d}. 
Remarkably, the correlations with $\delta(R)$ 
at $R>R_{\rm L}$ depend on halo mass much weaker 
than seen in Fig. \ref{fig_hpdcor}. To demonstrate this more clearly, 
we also show the results for other 12 mass bins, 
ranging from $10^{11.6}$ to $10^{14.4}\msun$ and with a bin width of 0.2 dex, as the gray 
lines. As one can see, the controlled correlations 
between halo properties and the over-density at $R>R_{\rm L}$ 
indeed have a weak dependence on halo mass.

These results clearly suggest that two distinct 
types of processes may regulate the halo properties 
simultaneously. The first is related to the mean over-density within the 
proto-halos, i.e. $\delta_{\rm L}$. 
According to the spherical collapse model, the mean linear over-density
should be the same for all halos identified at the same time.
The fact that $\delta_{\rm L}$ is different for different halos 
indicates that the spherical collapse model is not accurate. 
As shown in \cite{WangH07}, some halos need a higher 
over-density to form, because local tidal fields may act to 
prevent the collapse of the outer parts of proto-halos. 
This may explain the positive correlations of 
$\delta_{\rm L}$ with $\zf$ and $\vr$, 
and the negative correlation with $\lambda$. 
The second is the negative correlation of the over-density 
at $R>R_{\rm L}$ with both $\zf$ and $\vr$, and the positive 
correlation with $\lambda$, when $\delta_{\rm L}$ is 
controlled. As we will see later, these correlations  
are partly produced by the anti-correlation between 
the inner density of a proto-halo and the density 
exterior to $R_{\rm L}$. Both types of correlations 
have only weak dependence on halo mass, presumably because 
the linear density field is roughly scale-free over the scales 
concerned here. For convenience, we use internal correlations to refer 
to those related to $\delta_{\rm L}$, and external correlations 
to those related to $\delta_{\rm e}$ and over-densities on larger 
scales.

Since the three halo properties are correlated with each other,
it is important to check whether the internal and external 
correlations for a specific halo property are induced by the 
correlations for another halo property.
In Fig. \ref{fig_hpdiL}, we show the correlation coefficients
between $\delta_{\rm L}$ and one halo property with 
another property controlled. When $z_{\rm f}$ is fixed, 
the correlations with both $\vr$ and $\lambda$ are largely 
reduced and close to zero. In contrast, the correlation with $\zf$ 
remains strong even when $\vr$ or $\lambda$ is fixed. 
This suggests that the $\zf$-$\delta_{\rm L}$ correlation 
is the dominant one, while the other two correlations are largely  
the secondary effects of the $\zf$-$\delta_{\rm L}$ correlation.

In Fig. \ref{fig_partial}, we show partial external correlation 
coefficients as functions of $R$, with $\zf$ or $\vr$ or $\lambda$ controlled 
in addition to $\delta_{\rm L}$. We see that the external correlation 
for $\vr$ changes little when $\zf$ or $\lambda$ is controlled. 
Similarly, $\vr$ and $\zf$ have only 
little influence on the external correlation for $\lambda$. 
However, the external correlation for $\zf$ is strongly affected by 
the other two properties, in particular $\vr$.  When $\vr$ is
controlled, the external correlation efficiency of $\zf$
becomes close to zero over a large range of scales.
These results suggest that the external correlation of $\zf$ is largely 
the secondary effect of the external correlation of $\vr$, 
while the external correlation of $\lambda$ has a different origin 
from that of both $\zf$ and $\vr$.

\begin{figure}
    \centering
    \includegraphics[scale=0.6]{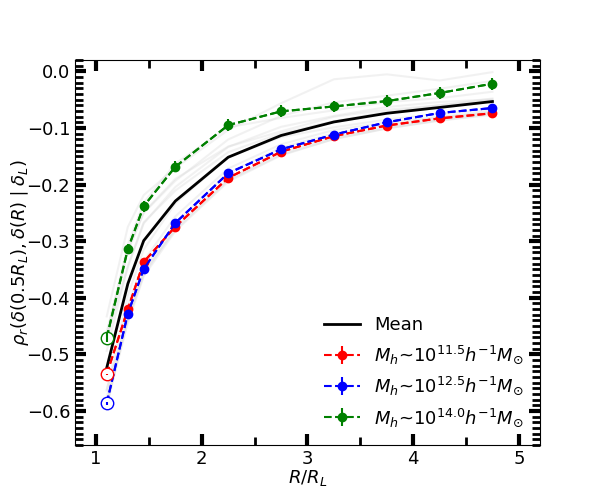}
	\caption{Partial correlation coefficient (Eq. \ref{eq_pcoe}) between  
	$\delta(0.5R_{\rm L})$ and $\delta(R)$, with $\delta_{\rm L}$ controlled, 
	as a function of $R/R_{\rm L}$. Here, $\delta(0.5R_{\rm L})$ is the average overdensity 
	within $[0.4-0.6]R_{\rm L}$. The symbols, color-code and error bars are the same as 
	in Fig. \ref{fig_partial}.}
	\label{fig_partial_in}
\end{figure}

To gain more understanding of the origin of the external correlations, 
we investigate the impact of the overdensity 
within $[0.4-0.6]R_{\rm L}$, denoted as $\delta(0.5R_{\rm L})$.
The partial correlation between $\delta(0.5R_{\rm L})$ 
and $\delta(R)$ with $\delta_{\rm L}$ 
controlled is shown in Fig. \ref{fig_partial_in}. 
It is similar to that of the external correlations of $\vr$ and $\zf$. 
At a given $\delta_{\rm L}$, $\delta(0.5R_{\rm L})$ 
measures the density slope within the proto-halo.
The anti-correlation between $\delta(0.5R_{\rm L})$ 
and the linear density field immediately exterior to $R_{\rm L}$
indicates that the density slope is coherent over a range of scales 
around $R_{\rm L}$. As suggested in \citet{Lu2006} and \citet{Dalal08}, the 
linear density profile of a proto-halo affects its accretion 
history and its final properties, such as $\zf$ and $\vr$, 
which may explain, at least partly, the external correlations of  $\vr$ and $\zf$.
The density slope may also affect $\lambda$, not only because it is correlated  
with the moment of inertial tensor of the proto-halo, which couples 
the proto-halo with the tidal field, but also because of 
its correlation with the halo assembly time over which the external 
tidal field operates to generate the angular momentum.

\section{On the origin of the secondary bias and its mass dependence}
\label{sec_orgsb}

Using the results obtained above, we attempt to understand the origin of the 
secondary bias and its mass dependence on.
From Eq. \ref{eq_pcoe}, we can write the Pearson 
coefficient between a halo property, $p$, and $\delta_{\rm b}$ as
\begin{equation}
    \rho_{\rm r}(p, \delta_{\rm b})=\alpha\rho_{\rm r}(p, \delta(r_{\rm b})|\delta_{\rm L})+
    \rho_{\rm r}(p, \delta_{\rm L})\rho_{\rm r}(\delta_{\rm b},\delta_{\rm L})\label{eq_dR}
\end{equation}
where $r_{\rm b}=R_{\rm b}/R_{\rm L}$, $\delta(r_{\rm b})$ is exactly $\delta_{\rm b}$, 
and
\begin{equation}
    \alpha=\sqrt{(1-\rho^2_{\rm r}(p, \delta_{\rm L}))(1-\rho^2_{\rm r}(\delta_{\rm b},\delta_{\rm L}))}\,.
\end{equation}
Since $\rho_{\rm r}(p, \delta_{\rm L})$ ranges 
from $\sim -0.3$ to $\sim 0.6$ and $\rho_{\rm r}(\delta_{\rm b},\delta_{\rm L})$ 
is about $0.1$, the value of $\alpha$ ranges from 0.8 to 1, 
with very weak dependence on halo mass. We thus can treat $\alpha$ as a constant
close to unity.

Eq. \ref{eq_dR} shows that the correlation of halo 
properties with $\delta_{\rm b}$ consists 
of two components, represented by the two terms 
on the right-hand side and  related to the external and 
internal correlations, respectively.  The secondary bias in the 
evolved density field, which is usually characterized by the 
$p$-$\Delta_{\rm b}$ correlation, may thus be understood 
in terms of the $p$-$\delta_{\rm b}$ correlation, 
as $\delta_{\rm b}$ is tightly correlated with 
$\Delta_{\rm b}$ and the correlation is quite independent 
of halo mass (Fig.\,\ref{fig_dbdib}).
The external correlation can result in a negative correlation 
of $\delta_{\rm b}$ with $\zf$ and $\vr$ and a 
positive correlation with $\lambda$. The internal correlation 
affects the secondary bias via the  
$\delta_{\rm L}$-$\delta_{\rm b}$ correlation, 
and can induce a positive correlation of 
of $\zf$ and $\vr$ with $\delta_{\rm b}$,
and a negative correlation between $\lambda$ and $\delta_{\rm b}$.
As shown in Figs. \ref{fig_hpdiL} and \ref{fig_partial}, 
both the external and internal components depend only 
weakly on halo mass. 
However, as we will show below, the two components combined 
can actually produce the mass dependence observed in the 
secondary bias.

Let us first look at the $\zf$ bias. For small halos with 
$\log(M_{\rm h}/\msun)=11.5$, the radius $R_b$ corresponds 
to $r_{\rm b}\equiv R_{\rm b}/R_{\rm L}>9.8$. 
At such a large scaled radius, the external correlation is 
very weak, with $\rho_{\rm r}(\zf, \delta(r_{\rm b})|\delta_{\rm L})$ 
close to zero (Fig. \ref{fig_partial}). In this case,  
$\rho_{\rm r}(\zf, \delta_{\rm b})\simeq\rho_{\rm r}(\zf, \delta_{\rm L})\rho_{\rm r}(\delta_{\rm b},\delta_{\rm L})$,
indicating that the positive correlation of $\zf$ with 
$\delta_{\rm b}$ for these halos is 
mainly determined by the internal correlation. 
For $M_*$ halos, where $r_{\rm b}=[4.56,6.84]$, 
the contribution from the external correlation is still negligible,
and so the dominant role is still played by the internal correlation.
Moreover, since the $\zf$-$\delta_{\rm L}$ and 
$\delta_{\rm b}$-$\delta_{\rm L}$ correlations 
for $M_*$ halos are similar to those for small halos
(Fig.\,\ref{fig_hpdcor} and \ref{fig_rhocor1}),
the contribution of the internal correlation 
for the two halo mass samples are also similar. Indeed, as shown in
Fig. \ref{fig_hpdcor}, the correlation coefficients 
of $\zf$ with $\delta_{\rm b}$ are very similar for 
the two halo samples.
Finally, for massive cluster-sized halos, 
the scaled radius $r_{\rm b}=[1.44,2.16]$. 
At such a scale, the external correlation becomes much stronger 
(Fig. \ref{fig_partial}), while the contribution of the internal correlation
is only slightly stronger than that for the other two halo mass bins. 
Consequently, the external correlation is able to counterbalance the 
internal correlation, making the secondary bias for these halos very weak.

Similar analysis can be made for $\vr$. 
The internal correlation of $\vr$ is weaker, while the external correlation 
is stronger, than that of $\zf$, making the external correlation more 
important for the $\vr$ bias. 
This also makes the mass-dependence of the $\vr$ bias different
from that of the $\zf$ bias. For small halos, 
the $\vr$ bias is still dominated by the internal correlation, because
their $\rho_{\rm r}(\vr, \delta(r_{\rm b})|\delta_{\rm L})$ 
is close to zero (Fig. \ref{fig_partial}).
Thus, a positive $\vr$-$\delta_{\rm b}$ correlation 
is expected, consistent with the results shown in 
Figs. \ref{fig_bias} and \ref{fig_vrdib}. 
For $M_*$ halos, the internal and external correlations become comparable, 
so that $\vr$ is independent of $\delta_{\rm b}$. 
For massive cluster-sized halos, where $r_{\rm b}$ 
becomes small, the external correlation becomes the 
dominant one,  producing a negative correlation between $\vr$ and 
$\delta_{\rm b}$. The change in the sign of the $\vr$ bias is 
thus a result of the competition between the two correlations. 

Finally for the spin, the internal correlation is even weaker, and 
thus plays only a minor role in producing the $\lambda$ bias. 
As shown in Fig. \ref{fig_partial}, the external correlation for the spin
decreases with $R/R_{\rm L}$ slowly, and so the dominant effect 
for the $\lambda$ bias is always the external correlation 
over the entire mass range. Since $r_{\rm b}=R_{\rm b}/R_{\rm L}$ 
decreases with increasing halo mass and the external correlation is 
stronger at smaller $r_{\rm b}$, an increasing trend 
with halo mass is produced, as shown in 
Figs. \ref{fig_bias}, \ref{fig_spdib} and \ref{fig_hpdcor}.

The analyses above suggest that the secondary bias 
observed in the evolved density field 
is primarily produced by the internal and external correlations
of halo properties with the linear density field, and that these 
correlations provide a transparent way to understand the origin of 
the secondary bias. 
We note that the results are not sensitive to the choice 
of $R_{\rm b}$, on which the halo bias is estimated. 
Indeed, we reached very similar conclusions using other scales, although 
the details may be different. Other factors, which are not 
included in our analysis, may also contribute to or affect the 
secondary bias. For example, both the external and internal 
correlations show weak but complicated dependence on halo mass, 
indicating that the shape of the linear power spectrum may also play 
a role. In addition, the correlations of the halo properties with 
$\delta_{\rm b}$ are not expected to be exactly the same as 
those with $\Delta_{\rm b}$, estimated at $z=0$, although these 
two overdensities are correlated. As shown in 
Figs. \ref{fig_zfdib}, \ref{fig_vrdib} and \ref{fig_spdib}, the 
non-linear evolution of the density field around halos
weakens the strength of the secondary bias slightly, 
presumably because of the mixing of scales of the linear
density field in the evolved field.

\section{Summary and discussion}\label{sec_sum}

In this paper, we use two $N$-body simulations, ELUCID and S1k, 
to study the origin of the secondary bias 
for halo assembly time ($\zf$), concentration ($\vr$) and spin 
($\lambda$), and its dependence on halo mass. 
Splashback halos, which have experienced 
strong non-linear processes, are excluded from our analyses. 
Our study is based on the correlations of these 
halo properties with linear densities at various scales. 
The main results can be summarized as follows.

\begin{itemize}
\item 
We find that the correlations of halo 
properties with the density measured at $R_{\rm b}=[10,15]\mpc$
are stronger in the linear density field than in the 
evolved density field at $z=0$. The secondary bias 
in the evolved density field is 
the secondary effect of the correlations of halo properties
with the linear density 
(Figs. \ref{fig_zfdib}, \ref{fig_vrdib}, \ref{fig_spdib} and \ref{fig_bias_cdib})
through the correlation of the two densities (Fig. \ref{fig_dbdib}). 
\item 
The correlations of halo properties with the linear density field
vary rapidly and even change signs around the halo Lagrangian radius
and approach zero at larger scales (Fig. \ref{fig_hpdcor}).
This suggests that the linear densities inside and outside the 
proto-halos affect the final halo properties in different ways.
\item 
The three halo properties are strongly correlated with
the average linear density within the Lagrangian radius, 
$\delta_{\rm L}$ (Fig. \ref{fig_hpdiL}). 
The $\zf$ and $\vr$ of halos increase with $\delta_{\rm L}$, 
while $\lambda$ exhibits a negative correlation with $\delta_{\rm L}$. 
The Pearson correlation coefficient shows that all the three 
correlations depend only weakly on the halo mass. 
Our results suggest that the internal correlations of $\vr$ and $\lambda$ 
are the secondary effects of the internal correlation of $\zf$.
\item 
At a given $\delta_{\rm L}$, all the three halo properties 
are correlated with $\delta(R/R_{\rm L})$ at $R>R_{\rm L}$ 
in a way opposite to the internal correlation (Figs. \ref{fig_hp2d} and \ref{fig_partial}). 
The conditional correlations, with $\delta_{\rm L}$ fixed, 
show only weak dependence on halo mass for all the three halo properties. 
The external correlations for $\zf$ and $\vr$ appear to share the same origin 
that may be partly related to the correlation of the inner slope of 
the proto-halos with the density field on large scales; 
the external correlation of $\lambda$ may have a different origin.
\item 
The internal and external correlations are the main drivers of 
the secondary bias (Eq. \ref{eq_dR}).
For the $\zf$ bias, the internal correlation dominates for 
almost the whole range of the halo mass, and the contributions of the 
two types of correlation are comparable for cluster-sized halos. For the 
$\vr$ bias, the internal correlation dominates for low-mass halos and 
the external correlation dominates for cluster-sized halos, 
while the two correlations are comparable for $M_*$ halos. 
For the $\lambda$ bias, the external correlation dominates over 
the whole mass range and the internal correlation only plays a minor role.
\end{itemize}

Our analysis provides a new perspective on the nature of the secondary bias. 
The results suggest that the combination of 
the internal and external correlations in the linear 
density field drives the secondary bias of the three halo properties 
considered here,  and that the mass dependence of the secondary bias 
is the result of the competition of the two correlations and the 
mass dependence of the contribution of the external correlation. 
We therefore need to understand the mechanisms that produce the internal 
and external correlations, in order to fully understand the 
origin of the secondary bias.

The internal correlation  
of $\zf$ has been studied in \cite{WangH07}, who 
found that the proto-halo of a $z=0$ halo with high $\zf$ 
is usually located in the vicinity of massive structures. 
They thus suggested that the large-scale tidal field
associated with (neighbour halos or/and large scale structures) the high density can truncate the  
accretion of the material in the outskirt of the proto-halo\citep[see also][]{Salcedo2018}. 
The stronger the tidal field is, the larger the truncation effect
and the higher the assembly redshift. This can explain 
the $\zf$-$\delta_{\rm L}$ correlation. 
However, they suggested that this effect is only important
for small halos \citep[see also e.g.][]{Dalal08, Mansfield20}, 
which is not consistent with our results that this effect is equally
important for cluster-sized halos. This indicates that the 
tidal truncation may also be important for massive halos.

The external correlation can produce a secondary bias effect 
that is totally opposite to the internal correlation. 
The external correlation is complicated and more than one mechanism may be 
relevant. For example, previous studies \citep[e.g.][]{Fakhouri09, WangH11} 
suggested that dense environments can enhance the frequency of halo merging, 
and mass accretion, producing an effect similar to that of the external correlation
we observe. The inner density profile within a proto-halo, 
which is found to correlate with large-scale density (Fig. \ref{fig_partial_in}),
can also affect the assembly history and concentration of the halo
\citep[e.g.][]{Lu2006, Dalal08}. A steeper initial profile, in general, produces an 
older and more concentrated final halo, leading to the external correlation. 
External tidal field can not only truncate mass accretion onto a halo 
but can also accelerate the tangential velocity of the material 
around the halo and enhance the spin \citep[e.g.][]{Shi15}.
This may explain the significant positive correlation between the spin and 
linear density at large scales. The steepness of the initial density profile,
which is correlated with the large-scale density field, can also affect 
the spin through its correlation with the moment of the inertial tensor of the 
proto-halo and its correlation with the  assembly time over which the external 
tidal field operates to generate the angular momentum. 

Clearly, further investigations are still needed to understand the details 
of these mechanisms and  how they affect the properties of halos. 
A detailed understanding of these mechanisms is essential not only 
for understanding the origin of the secondary bias, but also for 
constructing a comprehensive picture of halo formation in the cosmic density 
field. The results obtained here pave the way toward such a goal.

\section*{Acknowledgments}

We thank the anonymous referee for a useful report that significantly improves the paper. 
This work is supported by the National Key R\&D Program of China (grant No. 2018YFA0404503), 
the National Natural Science Foundation of China (NSFC, Nos.  11733004, 11890693, and 11421303), 
and the Fundamental Research Funds for the Central Universities. We acknowledge the 
science research grants from the China Manned Space Project with NO. CMS-CSST-2021-A03. 
The work is also supported by the Supercomputer Center of University of Science and Technology of China.

\bibliography{ref.bib}

\begin{thebibliography}{60}
\expandafter\ifx\csname natexlab\endcsname\relax\def\natexlab#1{#1}\fi

\bibitem[{{Allgood} {et~al.}(2006){Allgood}, {Flores}, {Primack}, {Kravtsov},
  {Wechsler}, {Faltenbacher}, \& {Bullock}}]{Allgood06}
{Allgood}, B., {Flores}, R.~A., {Primack}, J.~R., {et~al.} 2006, \mnras, 367,
  1781

\bibitem[{{Bett} {et~al.}(2007){Bett}, {Eke}, {Frenk}, {Jenkins}, {Helly}, \&
  {Navarro}}]{Bett07}
{Bett}, P., {Eke}, V., {Frenk}, C.~S., {et~al.} 2007, \mnras, 376, 215

\bibitem[{{Chen} {et~al.}(2016){Chen}, {Wang}, {Mo}, \& {Shi}}]{Chen16}
{Chen}, S., {Wang}, H., {Mo}, H.~J., \& {Shi}, J. 2016, \apj, 825, 49

\bibitem[{{Chen} {et~al.}(2020){Chen}, {Mo}, {Li}, {Wang}, {Yang}, {Zhang}, \&
  {Wang}}]{Chen20}
{Chen}, Y., {Mo}, H.~J., {Li}, C., {et~al.} 2020, \apj, 899, 81

\bibitem[{{Chue} {et~al.}(2018){Chue}, {Dalal}, \& {White}}]{Chue2018}
{Chue}, C. Y.~R., {Dalal}, N., \& {White}, M. 2018, \jcap, 2018, 012

\bibitem[{{Dalal} {et~al.}(2008){Dalal}, {White}, {Bond}, \&
  {Shirokov}}]{Dalal08}
{Dalal}, N., {White}, M., {Bond}, J.~R., \& {Shirokov}, A. 2008, \apj, 687, 12

\bibitem[{{Davis} {et~al.}(1985){Davis}, {Efstathiou}, {Frenk}, \&
  {White}}]{Davis85}
{Davis}, M., {Efstathiou}, G., {Frenk}, C.~S., \& {White}, S.~D.~M. 1985, \apj,
  292, 371

\bibitem[{{Desjacques}(2008)}]{Desjacques08}
{Desjacques}, V. 2008, \mnras, 388, 638

\bibitem[{{Dunkley} {et~al.}(2009){Dunkley}, {Komatsu}, {Nolta}, {Spergel},
  {Larson}, {Hinshaw}, {Page}, {Bennett}, {Gold}, {Jarosik}, {Weiland},
  {Halpern}, {Hill}, {Kogut}, {Limon}, {Meyer}, {Tucker}, {Wollack}, \&
  {Wright}}]{Dunkley09}
{Dunkley}, J., {Komatsu}, E., {Nolta}, M.~R., {et~al.} 2009, \apjs, 180, 306

\bibitem[{{Fakhouri} \& {Ma}(2009)}]{Fakhouri09}
{Fakhouri}, O. \& {Ma}, C.-P. 2009, \mnras, 394, 1825

\bibitem[{{Faltenbacher} \& {White}(2010)}]{Faltenbacher2010}
{Faltenbacher}, A. \& {White}, S. D.~M. 2010, \apj, 708, 469

\bibitem[{{Gao} {et~al.}(2005){Gao}, {Springel}, \& {White}}]{Gao05}
{Gao}, L., {Springel}, V., \& {White}, S. D.~M. 2005, \mnras, 363, L66

\bibitem[{{Gao} \& {White}(2007)}]{Gao07}
{Gao}, L. \& {White}, S. D.~M. 2007, \mnras, 377, L5

\bibitem[{{Gao} {et~al.}(2004){Gao}, {White}, {Jenkins}, {Stoehr}, \&
  {Springel}}]{Gao04}
{Gao}, L., {White}, S.~D.~M., {Jenkins}, A., {Stoehr}, F., \& {Springel}, V.
  2004, \mnras, 355, 819

\bibitem[{{Hahn} {et~al.}(2007){Hahn}, {Porciani}, {Carollo}, \&
  {Dekel}}]{Hahn07}
{Hahn}, O., {Porciani}, C., {Carollo}, C.~M., \& {Dekel}, A. 2007, \mnras, 375,
  489

\bibitem[{{Hahn} {et~al.}(2009){Hahn}, {Porciani}, {Dekel}, \&
  {Carollo}}]{Hahn09}
{Hahn}, O., {Porciani}, C., {Dekel}, A., \& {Carollo}, C.~M. 2009, \mnras, 398,
  1742

\bibitem[{{Han} {et~al.}(2019){Han}, {Li}, {Jing}, {Nishimichi}, {Wang}, \&
  {Jiang}}]{Han19}
{Han}, J., {Li}, Y., {Jing}, Y., {et~al.} 2019, \mnras, 482, 1900

\bibitem[{{Hearin} {et~al.}(2015){Hearin}, {Watson}, \& {van den
  Bosch}}]{Hearin15}
{Hearin}, A.~P., {Watson}, D.~F., \& {van den Bosch}, F.~C. 2015, \mnras, 452,
  1958

\bibitem[{{Jing} \& {Suto}(2002)}]{Jing02}
{Jing}, Y.~P. \& {Suto}, Y. 2002, \apj, 574, 538

\bibitem[{{Jing} {et~al.}(2007){Jing}, {Suto}, \& {Mo}}]{Jing07}
{Jing}, Y.~P., {Suto}, Y., \& {Mo}, H.~J. 2007, \apj, 657, 664

\bibitem[{{John} {et~al.}(1989){John}, {William}, \& Whitmore}]{John89}
{John}, N., {William}, W., \& Whitmore, G.~A. 1989

\bibitem[{{Johnson} {et~al.}(2019){Johnson}, {Maller}, {Berlind}, {Sinha}, \&
  {Holley-Bockelmann}}]{Johnson19}
{Johnson}, J.~W., {Maller}, A.~H., {Berlind}, A.~A., {Sinha}, M., \&
  {Holley-Bockelmann}, J.~K. 2019, \mnras, 486, 1156

\bibitem[{{Lacerna} \& {Padilla}(2011)}]{Lacerna11}
{Lacerna}, I. \& {Padilla}, N. 2011, \mnras, 412, 1283

\bibitem[{{Lazeyras} {et~al.}(2017){Lazeyras}, {Musso}, \&
  {Schmidt}}]{Lazeyras17}
{Lazeyras}, T., {Musso}, M., \& {Schmidt}, F. 2017, \jcap, 2017, 059

\bibitem[{{Li} {et~al.}(2013){Li}, {Gao}, {Xie}, \& {Guo}}]{LiR13}
{Li}, R., {Gao}, L., {Xie}, L., \& {Guo}, Q. 2013, \mnras, 435, 3592

\bibitem[{{Li} {et~al.}(2008){Li}, {Mo}, \& {Gao}}]{LiY08}
{Li}, Y., {Mo}, H.~J., \& {Gao}, L. 2008, \mnras, 389, 1419

\bibitem[{{Lu} {et~al.}(2006){Lu}, {Mo}, {Katz}, \& {Weinberg}}]{Lu2006}
{Lu}, Y., {Mo}, H.~J., {Katz}, N., \& {Weinberg}, M.~D. 2006, \mnras, 368, 1931

\bibitem[{{Ludlow} {et~al.}(2009){Ludlow}, {Navarro}, {Springel}, {Jenkins},
  {Frenk}, \& {Helmi}}]{Ludlow09}
{Ludlow}, A.~D., {Navarro}, J.~F., {Springel}, V., {et~al.} 2009, \apj, 692,
  931

\bibitem[{{Mansfield} \& {Kravtsov}(2020)}]{Mansfield20}
{Mansfield}, P. \& {Kravtsov}, A.~V. 2020, \mnras, 493, 4763

\bibitem[{{Mao} {et~al.}(2018){Mao}, {Zentner}, \& {Wechsler}}]{Mao18}
{Mao}, Y.-Y., {Zentner}, A.~R., \& {Wechsler}, R.~H. 2018, \mnras, 474, 5143

\bibitem[{{Mo} \& {White}(1996)}]{Mo1996}
{Mo}, H.~J. \& {White}, S.~D.~M. 1996, \mnras, 282, 347

\bibitem[{{Musso} \& {Sheth}(2012)}]{Musso12}
{Musso}, M. \& {Sheth}, R.~K. 2012, \mnras, 423, L102

\bibitem[{{Paranjape} {et~al.}(2018){Paranjape}, {Hahn}, \&
  {Sheth}}]{Paranjape2018}
{Paranjape}, A., {Hahn}, O., \& {Sheth}, R.~K. 2018, \mnras, 476, 3631

\bibitem[{{Ramakrishnan} {et~al.}(2019){Ramakrishnan}, {Paranjape}, {Hahn}, \&
  {Sheth}}]{Ramakrishnan19}
{Ramakrishnan}, S., {Paranjape}, A., {Hahn}, O., \& {Sheth}, R.~K. 2019,
  \mnras, 489, 2977

\bibitem[{{Salcedo} {et~al.}(2018){Salcedo}, {Maller}, {Berlind}, {Sinha},
  {McBride}, {Behroozi}, {Wechsler}, \& {Weinberg}}]{Salcedo2018}
{Salcedo}, A.~N., {Maller}, A.~H., {Berlind}, A.~A., {et~al.} 2018, \mnras,
  475, 4411

\bibitem[{{Salcedo} {et~al.}(2020){Salcedo}, {Zu}, {Zhang}, {Wang}, {Yang},
  {Wu}, {Jing}, {Mo}, \& {Weinberg}}]{Salcedo2020}
{Salcedo}, A.~N., {Zu}, Y., {Zhang}, Y., {et~al.} 2020, arXiv e-prints,
  arXiv:2010.04176

\bibitem[{{Sandvik} {et~al.}(2007){Sandvik}, {M{\"o}ller}, {Lee}, \&
  {White}}]{Sandvik07}
{Sandvik}, H.~B., {M{\"o}ller}, O., {Lee}, J., \& {White}, S.~D.~M. 2007,
  \mnras, 377, 234

\bibitem[{{Sheth} {et~al.}(2001){Sheth}, {Mo}, \& {Tormen}}]{Sheth2001}
{Sheth}, R.~K., {Mo}, H.~J., \& {Tormen}, G. 2001, \mnras, 323, 1

\bibitem[{{Shi} \& {Sheth}(2018)}]{shi18}
{Shi}, J. \& {Sheth}, R.~K. 2018, \mnras, 473, 2486

\bibitem[{{Shi} {et~al.}(2015){Shi}, {Wang}, \& {Mo}}]{Shi15}
{Shi}, J., {Wang}, H., \& {Mo}, H.~J. 2015, \apj, 807, 37

\bibitem[{{Springel}(2005)}]{Springel05}
{Springel}, V. 2005, \mnras, 364, 1105

\bibitem[{{Springel} {et~al.}(2001){Springel}, {White}, {Tormen}, \&
  {Kauffmann}}]{Springel01}
{Springel}, V., {White}, S. D.~M., {Tormen}, G., \& {Kauffmann}, G. 2001,
  \mnras, 328, 726

\bibitem[{{Tucci} {et~al.}(2021){Tucci}, {Montero-Dorta}, {Abramo},
  {Sato-Polito}, \& {Artale}}]{Tucci20}
{Tucci}, B., {Montero-Dorta}, A.~D., {Abramo}, L.~R., {Sato-Polito}, G., \&
  {Artale}, M.~C. 2021, \mnras, 500, 2777

\bibitem[{{Wang} {et~al.}(2018){Wang}, {Mo}, {Chen}, {Yang}, {Yang}, {Wang},
  {van den Bosch}, {Jing}, {Kang}, {Lin}, {Lim}, {Huang}, {Lu}, {Li}, {Cui},
  {Zhang}, {Tweed}, {Wei}, {Li}, \& {Shi}}]{WangH2018}
{Wang}, H., {Mo}, H.~J., {Chen}, S., {et~al.} 2018, \apj, 852, 31

\bibitem[{{Wang} {et~al.}(2009){Wang}, {Mo}, \& {Jing}}]{WangH09}
{Wang}, H., {Mo}, H.~J., \& {Jing}, Y.~P. 2009, \mnras, 396, 2249

\bibitem[{{Wang} {et~al.}(2011){Wang}, {Mo}, {Jing}, {Yang}, \&
  {Wang}}]{WangH11}
{Wang}, H., {Mo}, H.~J., {Jing}, Y.~P., {Yang}, X., \& {Wang}, Y. 2011, \mnras,
  413, 1973

\bibitem[{{Wang} {et~al.}(2016){Wang}, {Mo}, {Yang}, {Zhang}, {Shi}, {Jing},
  {Liu}, {Li}, {Kang}, \& {Gao}}]{WangH16}
{Wang}, H., {Mo}, H.~J., {Yang}, X., {et~al.} 2016, \apj, 831, 164

\bibitem[{{Wang} {et~al.}(2007){Wang}, {Mo}, \& {Jing}}]{WangH07}
{Wang}, H.~Y., {Mo}, H.~J., \& {Jing}, Y.~P. 2007, \mnras, 375, 633

\bibitem[{{Wang} \& {Kang}(2018)}]{WangP2018}
{Wang}, P. \& {Kang}, X. 2018, \mnras, 473, 1562

\bibitem[{{Wechsler} \& {Tinker}(2018)}]{Wechsler18}
{Wechsler}, R.~H. \& {Tinker}, J.~L. 2018, \araa, 56, 435

\bibitem[{{Wechsler} {et~al.}(2006){Wechsler}, {Zentner}, {Bullock},
  {Kravtsov}, \& {Allgood}}]{Wechsler06}
{Wechsler}, R.~H., {Zentner}, A.~R., {Bullock}, J.~S., {Kravtsov}, A.~V., \&
  {Allgood}, B. 2006, \apj, 652, 71

\bibitem[{{Wetzel} {et~al.}(2007){Wetzel}, {Cohn}, {White}, {Holz}, \&
  {Warren}}]{Wetzel07}
{Wetzel}, A.~R., {Cohn}, J.~D., {White}, M., {Holz}, D.~E., \& {Warren}, M.~S.
  2007, \apj, 656, 139

\bibitem[{{Xu} \& {Zheng}(2018)}]{xu18}
{Xu}, X. \& {Zheng}, Z. 2018, \mnras, 479, 1579

\bibitem[{{Yang} {et~al.}(2006){Yang}, {Mo}, \& {van den Bosch}}]{Yang06}
{Yang}, X., {Mo}, H.~J., \& {van den Bosch}, F.~C. 2006, \apjl, 638, L55

\bibitem[{{Yang} {et~al.}(2017){Yang}, {Zhang}, {Lu}, {Wang}, {Shi}, {Tweed},
  {Li}, {Luo}, {Lu}, \& {Yang}}]{Yang17}
{Yang}, X., {Zhang}, Y., {Lu}, T., {et~al.} 2017, \apj, 848, 60

\bibitem[{{Zel'Dovich}(1970)}]{ZelDovich70}
{Zel'Dovich}, Y.~B. 1970, \aap, 500, 13

\bibitem[{{Zentner}(2007)}]{Zentner07}
{Zentner}, A.~R. 2007, International Journal of Modern Physics D, 16, 763

\bibitem[{{Zentner} {et~al.}(2014){Zentner}, {Hearin}, \& {van den
  Bosch}}]{Zentner14}
{Zentner}, A.~R., {Hearin}, A.~P., \& {van den Bosch}, F.~C. 2014, \mnras, 443,
  3044

\bibitem[{{Zhao} {et~al.}(2003){Zhao}, {Jing}, {Mo}, \& {B{\"o}rner}}]{Zhao03}
{Zhao}, D.~H., {Jing}, Y.~P., {Mo}, H.~J., \& {B{\"o}rner}, G. 2003, \apjl,
  597, L9

\bibitem[{{Zhu} {et~al.}(2006){Zhu}, {Zheng}, {Lin}, {Jing}, {Kang}, \&
  {Gao}}]{Zhu06}
{Zhu}, G., {Zheng}, Z., {Lin}, W.~P., {et~al.} 2006, \apjl, 639, L5

\end{thebibliography}

\begin{appendix}
\section{The linear field at $z=18.4$ and its relation to the initial condition}

\begin{figure*}
    \centering
    \includegraphics[scale=0.35]{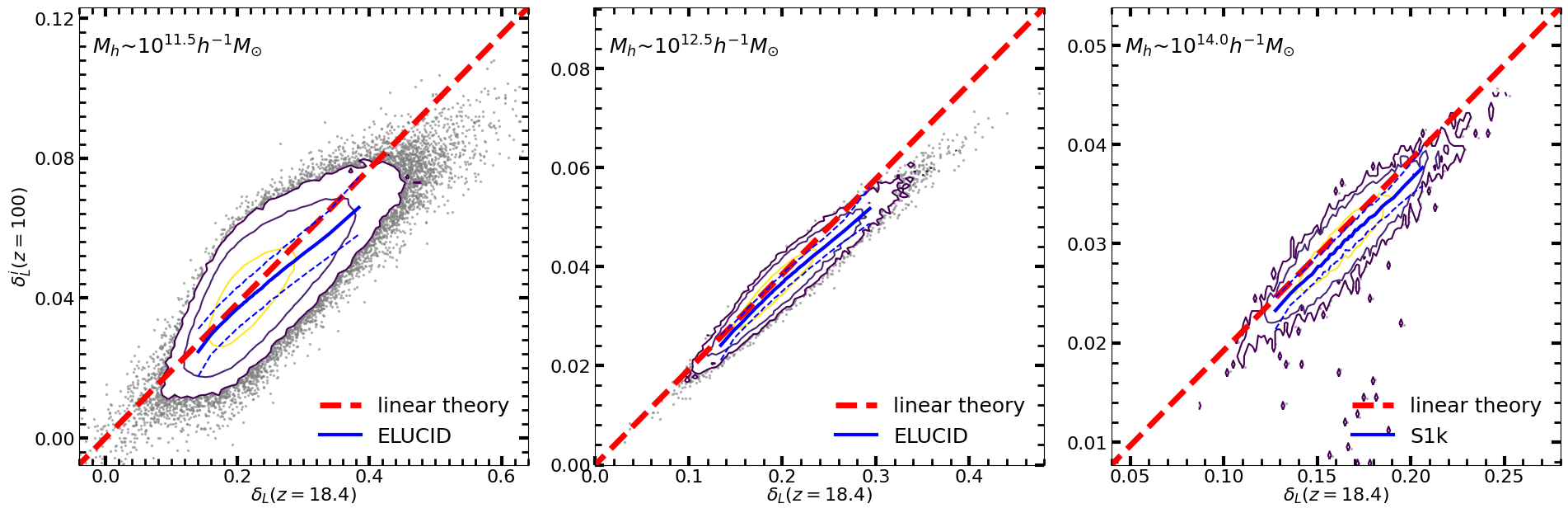}
	\caption{Contours show the correlation between $\delta^{i}_{\rm L}$ and $\delta_{\rm L}$ for three representative halo samples. The three  contour lines in each panel enclose 67\%, 95\% and 99\% of halos. The blue solid
	line shows the median value and the blue dash lines show 1$\sigma$ dispersion around the median relation. The red dash lines show the prediction of the linear theory.  }
	\label{fig_APP1}
\end{figure*}

We use the simulation snapshots at $z=18.4$, rather than the 
initial condition  (at $z=100$), to calculate the linear overdensities. 
Fig. \ref{fig_APP1} shows the comparison of the overdensities 
within the halo Lagrangian radius measured at $z=18.4$ with the initial 
condition for the three representative halo samples, together 
with the prediction of the linear theory. Here, we use 
$\delta_{\rm L}$ and $\delta^{\rm i}_{\rm L}$ to denote 
the overdensities at $z=18.4$ and $z=100$, respectively.
One can see that the median values of the three correlations
are consistent with the prediction of the linear perturbation theory, 
indicating that the density field at $z=18.4$ is still 
in the linear regime. These results justify the use  
of the simulation data at $z=18.4$ to estimate the linear densities.

\begin{figure*}
    \centering
    \includegraphics[scale=0.5]{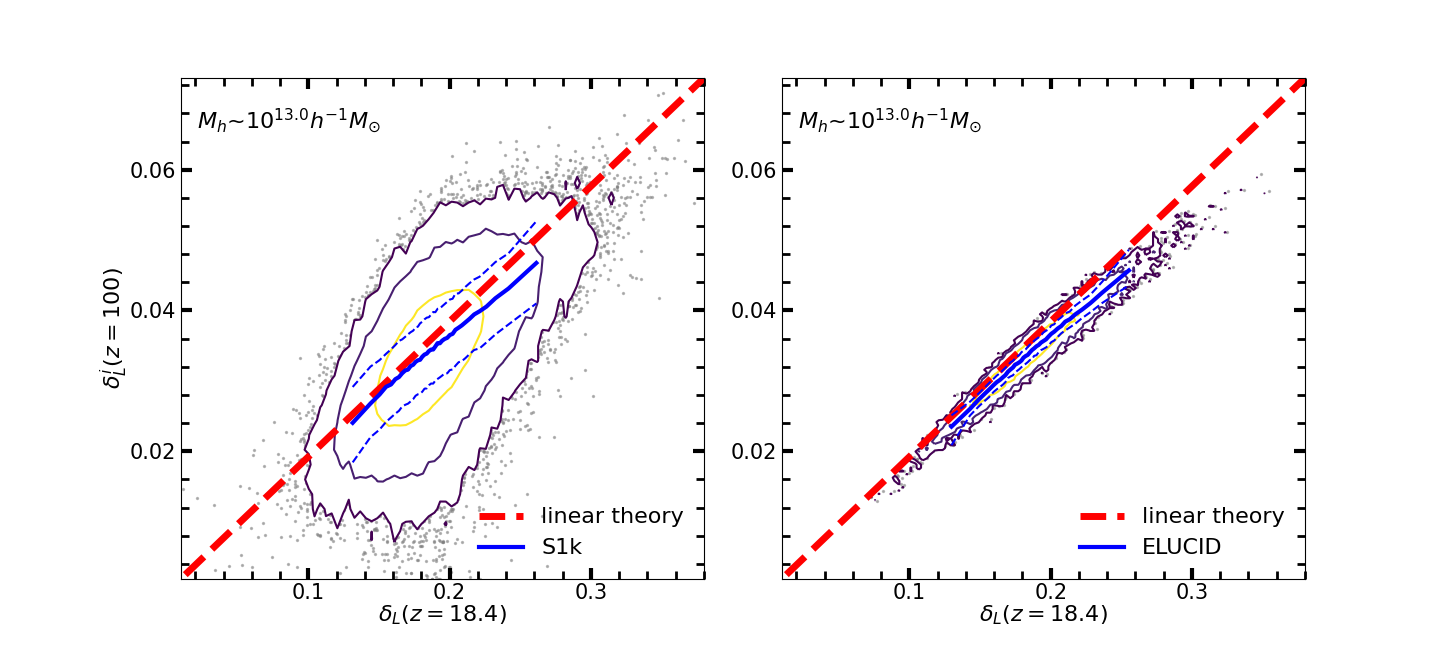}
	\caption{Similar to Fig. \ref{fig_APP1} but for halos with $\log M_{\rm h}=13.0$. The left and right panels show the results for S1k and ELUCID, respectively.}
	\label{fig_APP2}
\end{figure*}

When the mass resolution is not sufficiently high, 
there is a systematical problem in the initial condition 
due to shot noise. This problem  
is particularly severe for halos containing small number of particles.
As one can see, for $M_*$ and cluster-sized halos that contain 
at least 8,000 particles in each halo, the correlations
are very tight, while for the low-mass halos, each of which has 
only about 800 particles, the correlation is much poorer.
Thus, for halos with only about 800 particles, the
overdensities measured from the initial condition are severely affected
by the shot noise. To demonstrate this, we show the results for halos of
$\log(M_{\rm h}/\Msun)\sim 13$ of both simulations in Fig. \ref{fig_APP2}.
We choose this mass bin, because the S1k halos contain 
about 800 particles, similar to the low-mass halo sample. 
If shot noise were not important for these S1k halos, one 
would expect similar correlation between the two simulations. 
However, the correlation is much weaker for S1k than for ELUCID. 
Clearly,  shot noise can severely affect the measurements of the 
linear densities from the initial condition for halos containing 
less than 800 particles. We thus choose the snapshots at $z=18.4$
to calculate the linear overdensities rather than initial condition.

\end{appendix}
\bibliographystyle{aa}

\end{document}